\PassOptionsToPackage{square, comma, sort&compress}{natbib}
\documentclass[preprint,12pt]{elsarticle}

\usepackage{amssymb}
\usepackage{amsmath}
\usepackage{siunitx}
\usepackage[usenames,dvipsnames]{xcolor}
\usepackage{makecell}
\usepackage{gensymb}
\usepackage{geometry}
\usepackage{pdflscape}
\usepackage{chemformula}
\usepackage{tablefootnote}
\usepackage{mathtools}  % Used for braket symbol
\usepackage{braket}     % Used for braket symbol
\usepackage{threeparttable}
     
\usepackage[colorlinks=true,urlcolor=blue,linkcolor=blue,citecolor=blue]{hyperref}

\journal{Physics Reports}

\begin{document}

\begin{frontmatter}



\title{Remote Detection Optical Magnetometry}


\author[SUT]{Alexander M. Akulshin}
\author[UCB,JGU,HIM]{Dmitry Budker}
\author[ESO]{Felipe Pedreros Bustos}
\author[UP]{Tong Dang}
\author[JGU,HIM,UBFC]{Emmanuel Klinger}
\author[RS]{Simon M. Rochester}
\author[JGU,HIM]{Arne Wickenbrock}
\author[PKU]{Rui Zhang}

%affiliations%
\affiliation[SUT]{organization={Optical Sciences Centre, Swinburne University of Technology},
            %addressline={}, 
            city={Melbourne},
            postcode={3122}, 
            %state={California},
            country={Australia}}
            
\affiliation[UCB]{organization={Department of Physics, University of California},
            %addressline={}, 
            city={Berkeley},
            postcode={CA 94720}, 
            %state={California},
            country={USA}}
            
%\affiliation[ESO]{organization={European Southern Observatory},
%            %addressline={}, 
%            city={Garching b. M\"unchen},
%            postcode={85748}, 
%            %state={State Two},
%            country={Germany}}
            
\affiliation[UP]{organization={Department of Electrical and Systems Engineering, University of Pennsylvania},%Department and Organization
            %addressline={}, 
            city={Philadelphia},
            postcode={PA 19104}, 
            %state={State Two},
            country={USA}}
            
\address[JGU]{Johannes Gutenberg-Universit{\"a}t,  Mainz, 55128, Germany}

\address[HIM]{Helmholtz-Institut Mainz -- GSI Helmholtzzentrum f{\"u}r Schwerionenforschung, Mainz, 55128, Germany}

\address[ESO]{European Southern Observatory, Garching b. M\"unchen, 85748, Germany}

\address[UBFC]{Institut FEMTO-ST -- UMR 6174 CNRS, SupMicroTech-ENSMM, Universit{\'e} de Franche-Comt{\'e}, Besan{\c c}on, 25030, France}

%\affiliation[JGU]{organization={Johannes Gutenberg-Universit{\"a}t Mainz},%Department and Organization
%            %addressline={Staudingerweg 18}, 
%            city={Mainz},
%            postcode={55128}, 
%            %state={State Two},
%            country={Germany}}
%            
% \affiliation[HIM]{organization={Helmholtz-Institut Mainz, GSI Helmholtzzentrum f{\"u}r Schwerionenforschung},%Department and Organization
%        %addressline={Staudingerweg 18}, 
%        city={Mainz},
%        postcode={55128}, 
%        country={Germany}}
%        
%\affiliation[UBFC]{organization={Institut FEMTO-ST -- UMR 6174 CNRS, Universit\'e de Franche-Comt\'e, SupMicroTech-ENSMM},
%            city={Besancon},
%            postcode={25030}, 
%            country={France}}
%        
\affiliation[RS]{organization={Rochester Scientific LLC.},%Department and Organization
        %addressline={Staudingerweg 18}, 
        city={El Cerrito},
        postcode={CA 94530}, 
        country={USA}}
        
\affiliation[PKU]{organization={State Key Laboratory of Advanced Optical Communication Systems and Networks, Department of Electronics, and Center for Quantum Information Technology, Peking University},%Department and Organization
        %addressline={Staudingerweg 18}, 
        city={Beijing},
        postcode={100871}, 
        country={China}}
        
% \affiliation{, Beijing 100193, China}

\begin{abstract}
 Sensitive magnetometers have been applied in a wide range of research fields, including geophysical exploration, bio-magnetic field detection, ultralow-field nuclear magnetic resonance, etc. Commonly, magnetometers are directly placed at the position where the magnetic field is to be measured. However, in some situations, for example in near space or harsh environments, near nuclear reactors or particle accelerators, it is hard to place a magnetometer directly there. If the magnetic field can be detected remotely, i.e., via stand-off detection, this problem can be solved. As optical magnetometers are based on optical readout, they are naturally promising for stand-off detection. We review various approaches to optical stand-off magnetometry proposed and developed over the years, culminating in recent results on measuring magnetic fields in the mesosphere using laser guide stars, magnetometry with mirrorless-lasing readout, and proposals for satellite-assisted interrogation of atmospheric sodium.
\end{abstract}

% \includegraphics[width=1\textwidth]{GraphAbtract.pdf}



\end{frontmatter}

\tableofcontents

\clearpage
\section{Remote-detection magnetometry: motivations, applications, challenges}
\label{sec:intro}  % \label{} allows reference to this section
%·         why remote (stand-off detection)? -> Harsh environment; near space;  security,… (~1 paragraph)

\subsection{Overview and summary of ideas}

Sensitive magnetometers have been applied in a wide range of research fields, including  geophysical exploration~\cite{PARKER2010Geophysics,de2007spatial,pilipenko2017ulf}, biomagnetic-field detection~\cite{murzin2020ultrasensitive,Fu2020sensitive}, ultralow-field nuclear magnetic resonance~\cite{Bevilacqua2016,Tayler2017Invited}, urban science \cite{bowen2019network, dumont2022cities} etc. In addition to research, there are numerous applications of magnetometry in industry, security, and defense \cite{Fu2020sensitive}, including monitoring electrical-grid systems, detection and characterization of concealed/buried facilities such  as pipes and tunnels, detection of underground movement of personnel and equipment, etc.

Typically, magnetometers are placed directly at the position where the magnetic field is to be measured. However, in some situations, for example, in near-space~\cite{Higbie2011} or some harsh environments~\cite{Fu2020sensitive}, it is hard or impossible to place a magnetometer directly there. If the magnetic field can be detected remotely, that is, via stand-off detection, this problem can be solved. As optical magnetometers~\cite{budker2007optical} are based on optical readout, they naturally hold promise for stand-off detection; this approach has been explored in recent years.

For certain applications such as detection of magnetic anomalies of man-made origin (such as that due to submarines) or geophysical nature, remote magnetometry in the atmosphere is attractive for a number of reasons. First, the spatial range of the measurement is only limited by how far a laser beam can reach. In addition, the common problems of ``platform noise,'' the inevitable magnetic-field sources associated with the vehicle or the magnetometer-station infrastructure, are automatically avoided. If the measurement is performed sufficiently high above the surface, the local magnetic noise (e.g., due to power lines, localized thermoelectric currents, etc.) is also reduced.

%·         A brief review of earlier work (Rui can contribute here) (~1 paragraph)
The concept of stand-off magnetometry started in the 1970s~\cite{davis1989remas}, focusing on optical magnetometry based on atoms or molecules already present in the atmosphere, such as xenon~\cite{davis1989remas,Chng2017Remote} and molecular oxygen~\cite{davis1989remas,johnson2014remote}, see Sec.\,\ref{sec:REMAS}. 

There is also a proposal based on light reflection from the sea-water surface or a submerged object~\cite{epstein2016an}, see Sec.\,\ref{sec:SeaSurfaceReflectionMag}. There are two mechanisms that lead to polarization rotation: the magneto-optical Kerr effect upon reflection and the Faraday rotation in water. 

Another approach to remote magnetometry is to deploy a vapor cell and a retroreflector to the region where magnetic field is to be measured. 
Detection from a stand-off distance of up to 10\,m was demonstrated using light reflected from a mirror behind an alkali vapor cell~\cite{patton2012remotely}. Replacing the mirror with a true retroreflector could enable adjustment-free remote magnetometry, as discussed in Sec.\,\ref{sec:LightReflectionMag}.

More recently, magnetometry based on mesospheric sodium layer was proposed~\cite{Higbie2011}, modeled \cite{bustos2018simulations}, and experimentally demonstrated~\cite{Kane2018,Pedreros2018,Fan2019remote}. In these experiments, the light intensity or polarization \cite{Pedreros2018polarization} are modulated at the Larmor frequency (or its harmonic or subharmonic), leading to a change in the brightness of the sodium fluorescence, see Sec.\,\ref{sec:LGS-mag}.
%Polarization-driven spin precession of mesospheric sodium atoms~\cite{bustos2018polarization}

A current limitation of sky magnetometry is the small count of detected photons: the solid angle for collecting fluorescence from mesospheric sodium is typically less than $10^{-10}$. Another issue is the difficulty of operating in the presence of sunlight, limiting the operation to night-time hours. This limitation can be, at least in principle, overcome with mirrorless lasing (ML) techniques \cite{Zhang2021_Standoff} discussed in more detail in Sec.\,\ref{sec:mirrorless-lasing} or via detection of optical rotation in the mesospheric sodium layer of laser light launched from the ground with a detector on board a satellite \cite{dang2022satellite}, as discussed in Sec.\,\ref{sec:sat-mag}.

Finally, modern nonlinear optics techniques allow achieving remote lasing of the principal constituents of the atmosphere, which could also open prospects for remote magnetometry, as discussed in Sec.\,\ref{sec:Air_Lasing}.

\begin{figure}[htb]
    \centering
    \includegraphics[width=1\textwidth]{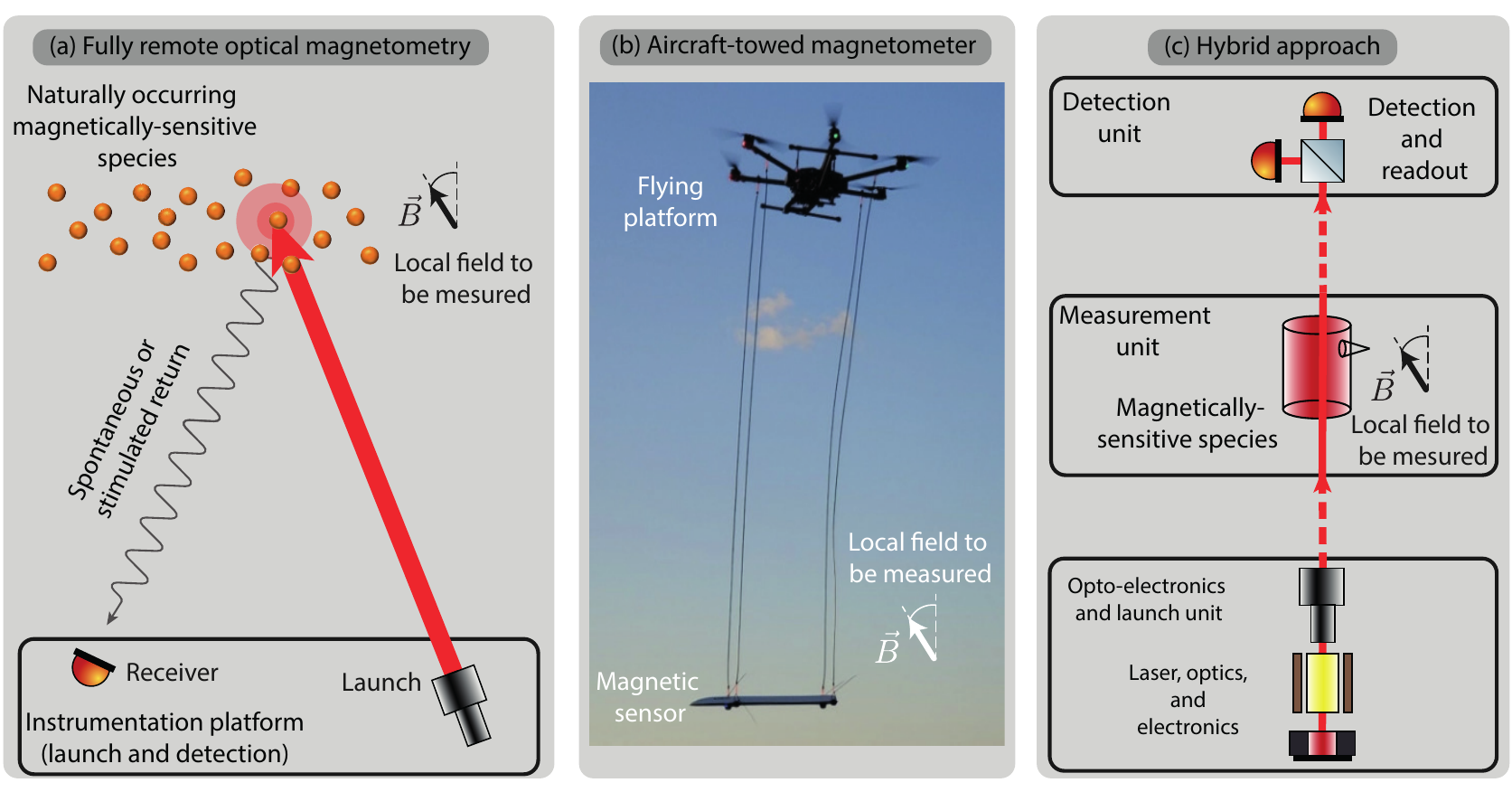}
    \caption{A summary of the different methods to perform remote magnetic field sensing to be reviewed. (a) Fully remote optical magnetometry: a laser beam illuminates naturally occurring species where the magnetic fields is to be measured. The detection is realized through spontaneous (Sec.\,\ref{sec:REMAS} and \ref{sec:LGS-mag}) or stimulated (Sec.\,\ref{sec:mirrorless-lasing} and \ref{sec:Air_Lasing}) return. (b) An optical magnetometer (here, MagArrow) is towed to an aircraft to perform magnetic survey. This approach, reviewed in detail in Ref.\,\cite{Zheng2021}, is enabled by the advent of compact, low power optical magnetometers, see Sec.\,\ref{sec:CPT-marnetometry}. Here, the magnetometer is held away from the drone to reduce the platform noise. Image adapted from \cite{zhang2022motion}. (c) Hybrid approach: the different units to realize the measurements are separated in space. Using a retro-reflector, the launch unit and detection unit can be placed at the same location (Sec.\,\ref{sec:LightReflectionMag}). The magnetically sensitive species can be naturally occurring (Sec.\,\ref{sec:sat-mag}) or contained, for example, in a vapor cell.
    }
    \label{fig:summary-of-schemes}
\end{figure}

% \DB{
A summary of the main approaches to remote magnetometry is presented in Fig.\,\ref{fig:summary-of-schemes}.
% }

%\DB{The intro is not yet finished; need to mention all sections}
%\subsection{Remotely interrogated Rb Magnetometer}
%\EK{A subsection on remote detection magnetometer  based on Optical magnetometry, Sec. %13.2 \cite{kimball2013optical}...}

\subsection{Synthetic gradiometry}

\begin{figure}
    \centering
   \includegraphics[width=\textwidth]{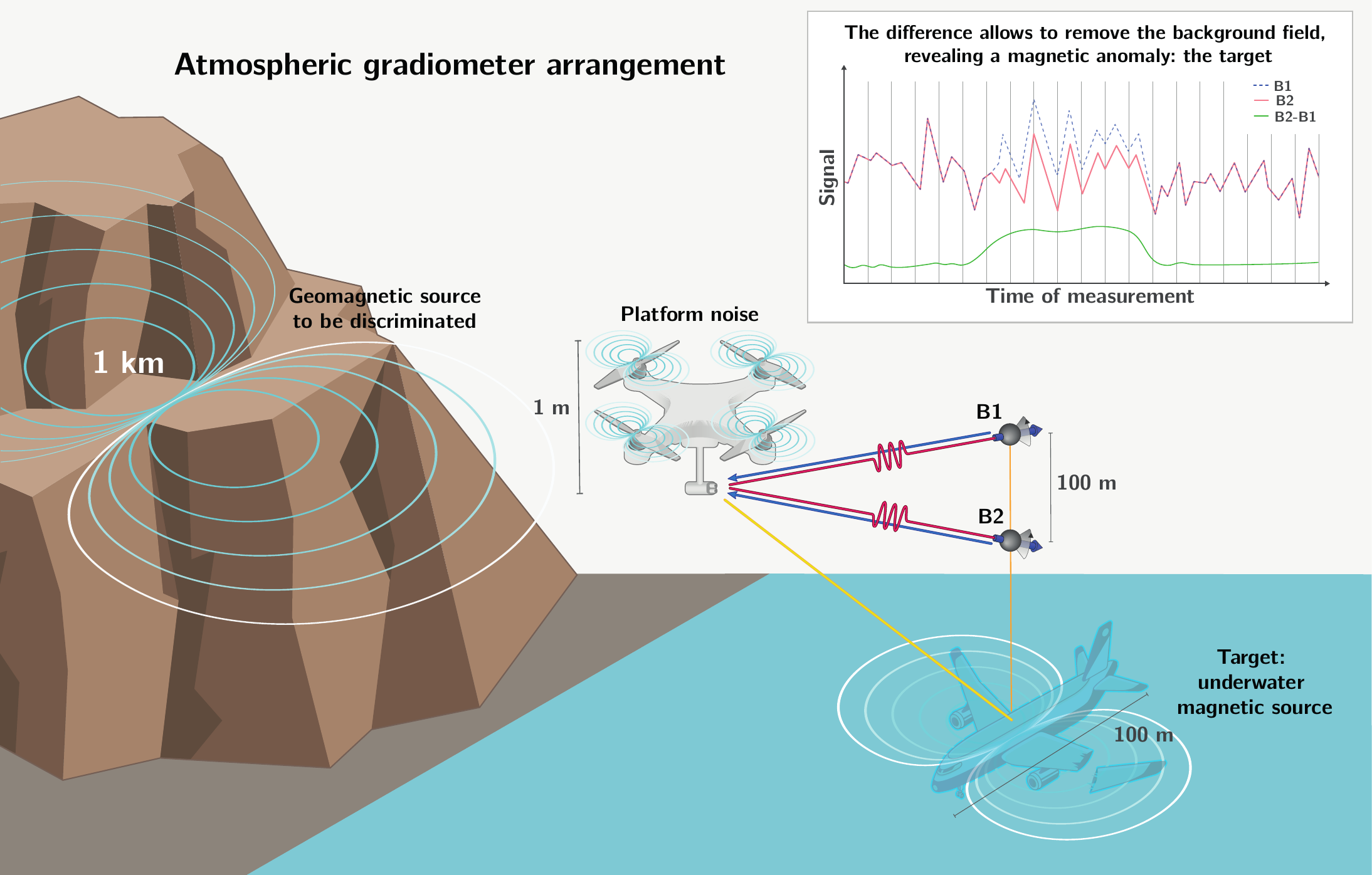}
    \caption{An application of synthetic gradiometry. Performing differential magnetic measurements over an adjustable baseline allows optimizing the baseline for the expected length scale of the target. This allows suppression of spurious signals from the platform (that have a characteristic scale of the platform size, typically much smaller than the target) and from geomagnetic sources (with spatial scales that are typically much larger than those of the target).}
    \label{Fig:Synthetic_Grad}
\end{figure}

With remote atmospheric magnetometry, we can make measurements at spatial locations of our choice. This can be useful in constructing magnetic-field maps, and in detection of magnetic anomalies. Specifically, one can construct a ``synthetic gradiometer'' and choose the distance between the sensors to maximally discriminate the sought-for signal from the different backgrounds. This concept is illustrated with a suggestive schematic in Fig.\,\ref{Fig:Synthetic_Grad}. Let us say, we would like to detect a submarine of characteristic dimensions of 100\,m using a small aircraft platform with characteristic dimensions of 1\,m. Let us say, that this is happening in an area with a typical dimension of geological structures (underwater canyons, mountains, etc.) of 1\,km. The anomaly detection needs to be done in the presence of significantly inhomogeneous and possibly noisy backgrounds from both the aircraft and the surrounding large-scale structures. Remote magnetometry allows to measure far away (say, hundreds of meters away from the platform), minimizing its fields that drop as $(r/R)^3$. The ability to choose the gradiometric base comparable to the characteristic fall-off distance of the target signal, in turn comparable to the dimensions of the source, is thus a powerful tool for discriminating the target and the spurious signals. 

We note that magnetic gradiometry measurements are also useful in applications beyond atmospheric magnetometry, for instance, in conjunction with target search and navigation of underwater vehicles \cite{naomiunderwater2009,hu2018multiple}.
%\EK{Check for potential reference on synthetic gradiometry, maybe these? \cite{naomiunderwater2009,hu2018multiple}}

\section{Earlier work} \label{sec:early-work}

We are not aware of previous dedicated reviews of remote detection optical magnetometry apart from the chapter by S.\,M.\,Rochester~\textit{et al} in Ref.\,\cite{kimball2013optical}. The key idea discussed there is that optical magnetometry is ideally suited for remote measurements because the atoms or molecules being interrogated, apart from the magnetic field being measured, only interact with optical fields that can be delivered with laser beams and observed, for example, using telescopes. Since sensitive magnetometry (as any other sensitive spectroscopic technique) typically requires parameter modulation of some sort, intensity-, frequency-, or polarization-modulated light fields are used. Such all-optical magnetometry techniques using modulation light have roots in the pioneering work by Bell and Bloom going back to mid 20th century \cite{bell1961optically}. In the rest of this Section, we trace the development of the ideas of remote optical magnetometry.

\subsection{The REMAS project} \label{sec:REMAS}

The original remote magnetic sensing (REMAS) work by William Happer and others, summarized in a comprehensive report \cite{davis1989remas}, was motivated by the need of detecting magnetic anomalies of natural and man-made nature (such as a submarine) in situations where locating the sensor in the vicinity of the anomaly is either impossible or impractical. The key idea was to measure magnetic fields using species already present in the atmosphere. Since magnetic detection is ultimately based on the Zeeman effect, the species need to be paramagnetic, if not in the ground state, then, at least, in an excited state. The general idea of the measurement is as follows. A laser beam is launched from a remote apparatus, for example, on board an aircraft or a surface vehicle, which propagates to the measurement region and excites the probe species, creating transient spin polarization. The spins evolve in the presence of the total magnetic field (including the anomaly). This precession is then detected by observing optical signal at the remote apparatus. An example of the signal is laser-induced fluorescence, affected by the magnetic field via the Hanle effect (see, for example, \cite{Budker2002RMP} and references therein).

The choice of the species to consider for remote sensing is influenced by a variety of factors, including the abundance in the atmosphere, the existence of suitable transitions for laser excitation and detection, the magnetic properties of the levels involved, etc. One specific example of a deleterious effect is fast  collisional relaxation in the presence of the atmosphere that broadens magnetic resonance and diminishes the effect of the magnetic field one seeks to measure.
In the end, the REMAS study found overwhelming difficulties with each of the species considered and came to a conclusion that remote atmospheric magnetometry may only become possible with radical breakthroughs in the technology. We recap the main conclusions of the REMAS study in Sec.\,\ref{subsec:REMAS_conclusions} below.    

Let us briefly summarize the species considered in the REMAS project. As a preliminary remark, we remind the reader of some of the units of magnetic field used in the literature (with units of magnetic field and flux density often not distinguished for nonmagnetic media). The Gaussian unit is gauss (G), $1\,\text{G}=10^{-4}$\,T, and a unit commonly used in geophysics is gamma, $1\,\text{gamma}=10^{-5}\,\text{G}=1$\,nT.

\subsubsection{Molecular oxygen}

Oxygen, constituting about 20\% of atmospheric molecules (being second only to nitrogen, N$_2$) and having a (rare for dimers) paramagnetic ground $^3\Sigma^-_g$ state (with the magnetic moment of 2.0\,$\mu_B$, $\mu_B$ being the Bohr magneton) and a convenient optical transition at 762\,nm, is a natural candidate for remote sensing. However, even in this case, REMAS reached a pessimistic overall conclusion due to the short (on the order of nanoseconds) decay time of spin polarization and significant line broadening due to collisions exacerbated by relatively small excitation cross-section and emission probability.

\begin{figure}[htb]
    \centering
\includegraphics[width=0.8\textwidth]{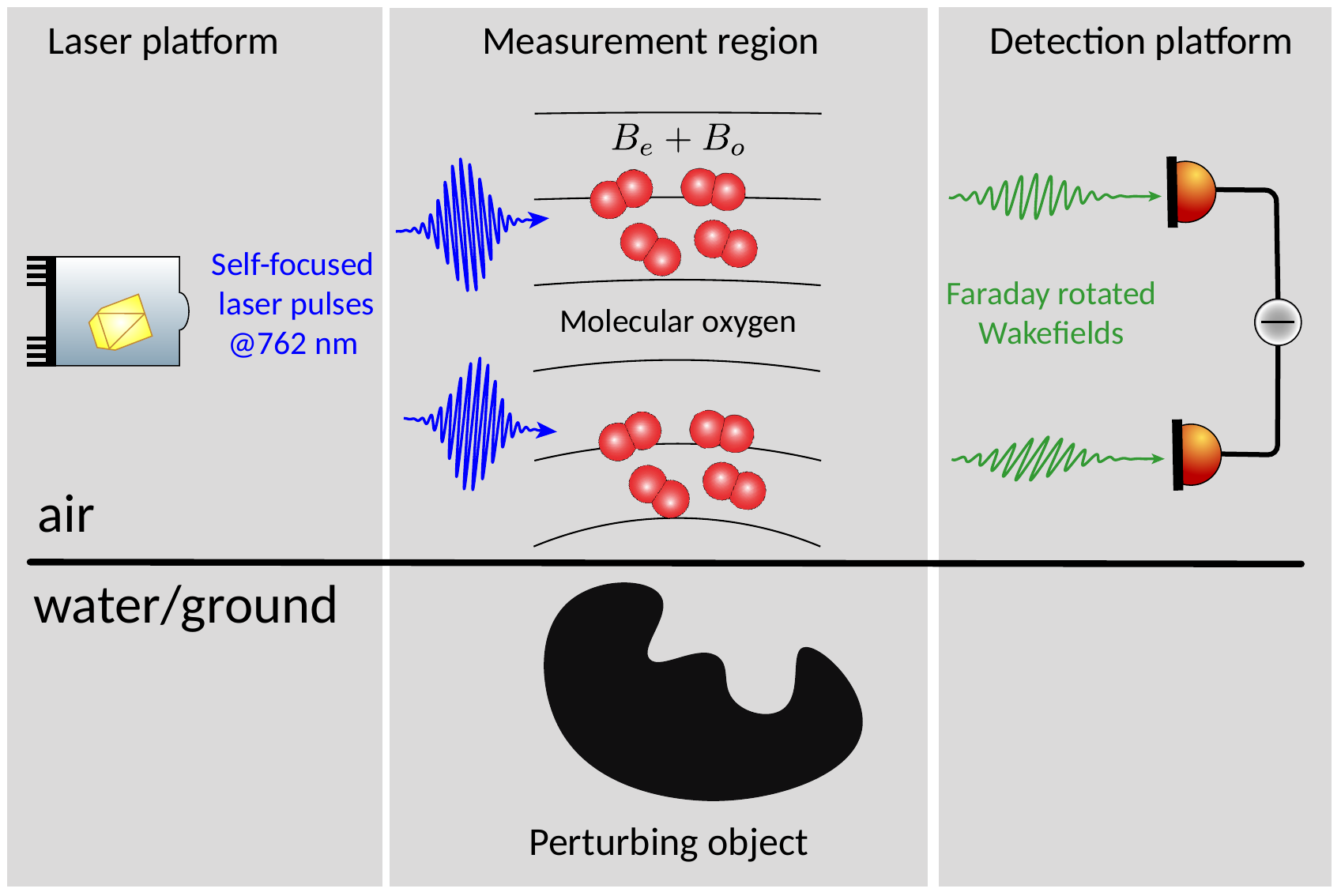}
    \caption{Schematic drawing of the remote atmospheric magnetometry technique proposed in \cite{johnson2014remote}. Light emitted from the laser platform is focused at a desired position, the measurement region. A laser pulse induces an electromagnetic wakefield behind the laser pulse, the polarization of which rotates in the background magnetic field. If an object is perturbing the local magnetic field, a difference in polarization rotation can be detected using a polarimetric gradiometer.}
    \label{fig:remote-oxygen}
\end{figure}

More recently, the concept of remote atmospheric magnetometry was revisited from a more modern perspective \cite{johnson2014remote}. As in the original REMAS project, the authors considered the $b^1\Sigma^+_g - X^3\Sigma^{-}_g$ magnetic dipole transition band of oxygen near 762\,nm. The novel aspects of the work include self-focusing of a laser pulse at the remote location where the magnetic field is to be measured and the use of the ``magnetic wakefield'' trailing the laser pulse, where the magnetic field polarization depends on the ambient field. The concept of the method is depicted in Fig.\,\ref{fig:remote-oxygen}.

\subsubsection{Atomic xenon}
The next species studied within REMAS was atomic $^{129}$Xe. Here, the ground-state magnetism is purely nuclear (the magnetic moment is approximately $-0.78\,\mu_N$, $\mu_N$ being the nuclear magneton), which ensures that the ground-state spin polarization is less sensitive to atomic collisions (longer relaxation times). 

On the other hand, a downside of using xenon is the low natural abundance of $^{129}$Xe, constituting 26.4\% of all atmospheric xenon with the atmospheric abundance of 87\,ppb for all Xe isotopes. An additional serious problem is the absence of optical transitions from the ground state, even for two-photon excitation.

\begin{figure}
    \centering
   \includegraphics[width=0.5\textwidth]{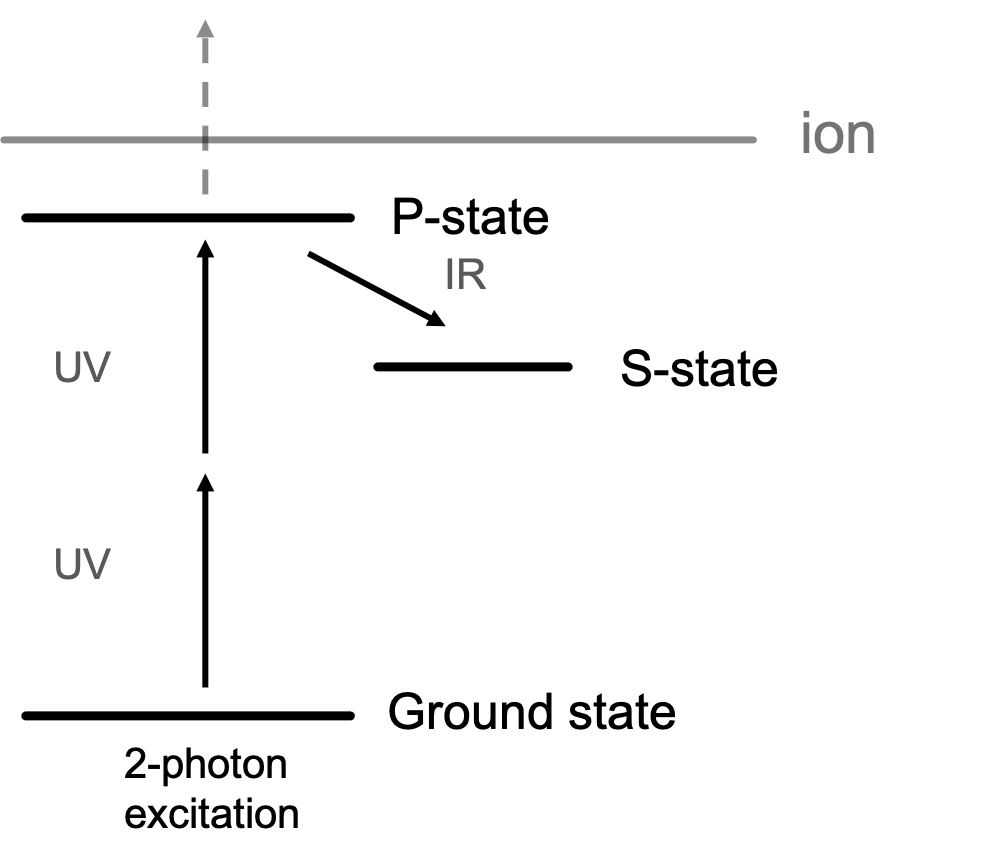}
    \caption{Xe energy levels and transitions considered in the REMAS project. See also Table\,\ref{Tab:Xe}.}
    \label{fig:REMAS_Xe}
\end{figure}

A schematic of the levels and transitions in Xe considered in REMAS is shown in Fig.\,\ref{fig:REMAS_Xe}. One of the $6p,\,J=2$, electronic states is excited via a two-photon transition driven by UV laser light. Absorption of the UV light in the atmosphere is an important consideration as it can be significant at the relevant wavelengths (see Table\,\ref{Tab:Xe}). Polarized pump laser light optically pumps the $^{129}$Xe nuclei in the cycle of emission and subsequent deexcitation of the upper state, which is predominantly nonradiatively quenched in collisions with the atmospheric gas or decays with emission of infrared (IR) light at wavelengths of around 830\,nm, the latter process important for detection.  The ground-state nuclear polarization is relatively long-lived. It relaxes, primarily due to collisions with atmospheric oxygen, with a characteristic decay time of 30\,s \cite{davis1989remas}. In practice, the effective relaxation time can be much faster due to effects such as diffusion and atmospheric wind. Consider, for example, a volume of a typical dimension of 1\,cm. There are always winds in the atmosphere with characteristic speed during even the nominally calm time of at least 1\,m/s. Thus, effective relaxation occurs on the scale faster that tens of milliseconds. 

In order to achieve optical pumping, it is essential for the pump light to predominantly excite one of the two upper hyperfine states.\footnote{In the case of radiative decay, it would not be necessary to resolve the hyperfine structure to accomplish optical pumping (see, for example, Prob.\,3.20 in \cite{budker2008atomic}. However, in this case, collisional redistribution in the excited state leads to the atoms returning to the two ground-state Zeeman sublevels with equal probability, which prevents polarization.} The same is also necessary to accomplish optical probing. The factor that hinders resolving the hyperfine states is pressure broadening that is comparable to the hyperfine intervals. The efficiency of optical pumping was characterized by the parameter
\begin{equation}
\label{eq:eta}
    \eta = \frac{R_{+}-R_{-}}{R_{+}+R_{-}}\,,
\end{equation}
where $R_{+}$, $R_{-}$ are rates of excitation of the spin up and spin down ground states, respectively.
$\eta$ depends on hyperfine splitting, pressure broadened width and laser linewidth. The values of $\eta$ estimated for the three possible two-photon transitions are presented in Table\,\ref{Tab:Xe}.

At low laser-light powers, the amount of polarized xenon grows proportionally to the laser-light power. However, this growth is limited at higher light intensities (higher than on the order of 1\,GW/cm$^2$) due to the competing process of two-photon-resonant three-photon ionization.

\begin{table}[htp]
\centering
\def\arraystretch{1.5}
\begin{tabular}{ccccccc}
\hline\hline
\makecell{Excited \\State} & \makecell{Energy \\(cm$^{-1}$)}& $\lambda$\,(nm) & \makecell{Absorption\\length} &\makecell{Hyperfine splitting\\ $v_{\text{hfs}}(5/2 - 3/2)$ } & \makecell {Pumping \\Efficiency, $\eta$}  \\
\hline
$6p{[}5/2{]}_{2}$ & 78120 & 256.0 & 455\,m %(8.9\,km)  
& 3.4\,GHz& 0.16    \\
$6p{[}3/2{]}_{2}$ & 79212 & 252.5 & 423\,m & 2.2\,GHz  & 0.11 \\
$6p'{[}3/2{]}_{2}$ & 89162 & 224.3 & 149\,m %(282\,m) 
& 7.2\,GHz & 0.31 \\
\hline\hline
\end{tabular}
\caption{Two-photon pumping of Xe in air. Absorption length of the UV light (defined here as the length over which the light intensity is attenuated by a factor of $e$) is an important factor that influences the range of a possible measurement. The values shown in the table are extracted from \cite{Strongylis1979absolute}, where the horizontal attenuation of UV light is given summing the contributions of absorption (notably by O$_2$ and O$_3$) and of Rayleigh scattering. The hyperfine splitting is between the two states in the upper $J=2$ P state with total angular momenta $F=5/2$ and 3/2. Pumping efficiency [Eq.\,\eqref{eq:eta}] characterizes how well the hyperfine components can be spectrally resolved. 
%\DB{Dima wonders: Felipe, can it really be that ESO does not have data for atmospheric transmission?}
}
\label{Tab:Xe}
\end{table}

%\EK{Air diffusion and WIND of Xe, see slide form J. Davis \cite{chen1999spatially}}

\subsubsection{Atomic krypton and neon}

Another possible candidate is $^{83}$Kr, with a natural abundance in the atmosphere of about 0.12\,ppm (about six time that of $^{129}$Xe), nuclear spin of $I=9/2$ and magnetic moment of about -0.97\,\,$\mu_N$. Note that, since the Larmor frequency goes as magnetic moment over the spin $I$, the Larmor frequency for $^{83}$Kr is about a factor of seven lower than that of $^{129}$Xe. A disadvantage of krypton compared to xenon is that the longest photon wavelength for two-photon excitation is 217\,nm, and the absorption length is on the order of 100\,m, %around \EK{178\,m (115 m)}
dominated by absorption by atmospheric oxygen. While the use of krypton for REMAS was not excluded in the study, no strong endorsement was given to this system either.  

Finally, neon was also considered and largely dismissed on the grounds of absorption length in the atmosphere for the excitation photons (133\,nm) is only 2$\cdot 10^{-4}$\,m, again due to oxygen.
%\TD{ref. p36 REMAS final report}

\subsection{Conclusions of the REMAS project}
\label{subsec:REMAS_conclusions} 

The main conclusions of the REMAS project \cite{davis1989remas} can be summarized as follows.
\begin{itemize}
    \item There is no clear path towards magnetometry based on directional light emission from the molecules or noble gases present in the atmosphere. We note here that alkali atoms present in the upper atmosphere (i.e., the part of the atmosphere above $\approx$\,50\,km) apparently were not considered.
    \item Magnetometry based on these species that does not rely on directional emission may still be possible. We are not aware of the REMAS schemes having been realized experimentally. However, applications to magnetic geological prospecting, mapping of underground facilities, or monitoring of fields from buried AC power lines did not appear plausible. Due to low expected signal-to-noise ratio, only applications allowing for long (hour-scale) times would be possible. In principle, this could be useful for monitoring change in magnetic fields over these time scales, but the applicability would be limited. 
    \item Technologies necessary for implementing REMAS in its limited form were found to be in various degrees of readiness. The absence of the required high pulse and average power UV lasers appeared to be a bottleneck, while infrared detectors and beam-pointing optics were largely available.
\end{itemize}

\subsection{Magnetometry based on light reflection from sea-water surface} \label{sec:SeaSurfaceReflectionMag}

\begin{figure}
    \centering
   \includegraphics[width=\textwidth/2]{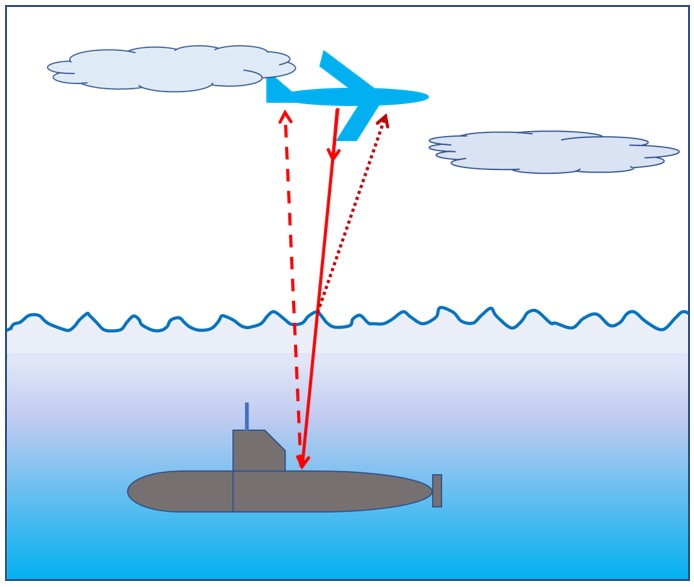}
    \caption{Scheme of remote magnetometry using reflection from water surface and underwater objects.}
    \label{Fig:Underwater objects}
\end{figure}

The polarization of the laser light reflected  from a remote object can carry information about the local magnetic field. If reflection occurs from sea water, this approach may be applied in a wide range of important applications including geological exploration and detection of underwater objects, e.g., submarines, perturbing the local ambient magnetic field, as shown in Fig.\,\ref{Fig:Underwater objects}.

The feasibility of this approach of remote magnetometry and, in particular, remote detection of underwater objects was considered in Ref.\,\cite{epstein2016an}.
The polarization rotation of light upon reflection is due to the surface magneto-optical Kerr effect.  In addition, light reflected from a submerged object might also contribute to the polarization rotation. In this case the polarization rotation is caused by the Faraday effect.  For both effects, the polarization rotation is proportional to the local magnetic field, but rotation due to the Faraday effect also depends on the propagation distance. 

A model describing both contributions to polarization rotation was developed in Ref.\,\cite{epstein2016an}, where nonlinear effects in water were neglected.
First a simplified configuration was considered in which a linearly polarized monochromatic laser beam propagates parallel to the Earth’s magnetic field, which is perpendicular to the surface of the water. A reflective underwater object located at a certain depth $h$ is represented by a perfectly conducting plate parallel to the surface. It was concluded that the polarization rotation due to the surface magneto-optical Kerr effect is small, on the order of $10^{-12}$\,rad. An analytical expression was obtained for the polarization-rotated field when the incident plane wave is at an arbitrary angle and polarization with respect to the water surface. It was found that no choice of polarization or angle of incidence significantly increases the Kerr-effect contribution to polarization rotation. Thus, it would be extremely challenging to use this effect for remote magnetometry.
The angle of rotation due to the Faraday effect is greater by a factor of roughly $h/\lambda \gg 1$, where $\lambda$ is the wavelength of the light. In the range of depth of $5<h<30$\,m, with a typical geomagnetic field, the rotation angle can be on the order of $10^{-3}$\,rad. While this is an easily detected polarization-rotation angle, measurements can be complicated by several factors such as the shape and conductivity of the underwater object. It was also noted that ripples of the seawater surface and bubbles might significantly attenuate the reflected light.

For effective detection of reflected light, the applied probe radiation must be sufficiently intense. Propagation of intense light in dispersive nonlinear media such as air or water is an important general problem since powerful continuous wave and intense pulsed laser radiation is widely used for various types of remote sensing and generation of beacons (guide stars).
 
 The influence of laser intensity and phase noise on the spatial and temporal evolution of laser radiation was considered theoretically in Ref.\,\cite{Isaacs2019JOSA}. As frequency noise can affect intensity noise and vice versa, it was shown that the Kerr effect can reduce the intensity noise under certain conditions. The developed model was used to study the transverse and longitudinal intensity instabilities. It was shown that significant spectral modification can occur if the initial intensity-noise level is sufficiently high. The authors considered the spatial and temporal evolution of laser radiation during propagation in the atmosphere at wavelengths of 0.85\,\textmu m, 1\,\textmu m and 10.6\,\textmu m.

The interaction of intense laser pulses with atmospheric gases is studied in the contexts of generation of broadband terahertz radiation via two-color photoionization currents in nitrogen, and  the generation of an electromagnetic wakefield by the induced magnetization currents of oxygen Ref.\,\cite{johnson2014PhD}. A laser pulse propagating in the atmosphere can excite magnetic
dipole transitions in molecular oxygen. The resulting transient current creates
a co-propagating electromagnetic field behind the laser pulse, i.e. the wakefield,
which has a rotated polarization with respect to that of the incident laser pulses, which depends on the background magnetic field.

%S. Sahoo et al, "Nonlinear magnetoelectric effect in atomic vapor and its application to precisionradio-frequency magnetometry", PRA 105, 063509 (2022).
%

%We demonstrate the nonlinear magnetoelectric (ME) effect in atomic vapor achieved through the parametric interaction of optical electric and radio-frequency (rf) magnetic fields leading to the generation of new optical electric fields. Density matrix calculations are performed to validate the experimental results. Moreover, the predicted dependence of the generated optical electric field amplitudes on the rf magnetic field strength is experimentally verified to confirm theME effect. The system provides a technique for precision rf magnetometry based on this phenomenon. We could experimentally achieve an rf magnetic field sensitivity of 70 fT/√Hz at 1 kHz to 7.5 pT/√Hz at 3 MHz for zero bias field in an unshielded environment. The appealing features of the proposed rf magnetometer using our system include a high dynamic range up to 1012, 6 dB bandwidth of450 kHz, and arbitrary frequency resolution, which are intrinsic to the nonlinear ME effect in the medium.
 
 %\clearpage
 \section{Remotely interrogated optical alkali-vapor magnetometers} \label{sec:LightReflectionMag}
 
 The ``brute-force'' remote magnetometry may consist in placing a part of or even complete magnetic sensor where the field needs to be measured. The readout of the sensor may be achieved optically or via radio telecommunication.  While not being ``remote'' in the strict meaning of the term, this approach may, in fact, be practical for certain applications, in particular in view of the rapid development of unmanned aerial (UAV) and undersea (UUV) vehicles, discussed in more details in Ref.\,\cite{Zheng2021}, as well as techniques for ``broadcasting'' low-cost sensors on the ground, water surface, or deploying them on air balloons or parachutes. Thus the developments in this direction involve finding ways to reduce the complexity, price, weight, and power consumption of the sensors. We briefly review such developments here.
 
 %The idea here is that part of or the complete magnetic field sensor can be placed in the region of interest. 
% \EK{Maybe the section name needs a small rework. We need to write a small introduction to say that these next subsections have essentially the same physics (OPM and CPT).} \AAk{I think it's more interesting to note the distinctive features of the underlying physics.}
 
\subsection{Magnetometer based on polarization rotation} \label{sec:RemotIntero-PolRotation}

In their pioneering work \cite{bell1961optically}, Bell and Bloom  realized that, for atoms, the effect of microwave fields used in optically-pumped magnetometers (OPM) could be achieved with modulated light. Although their work received major attention in the atomic physics community, see, for example, the review  \cite{alexandrov2005dynamic}, Earth-field magnetometers continued to rely on radio-frequency excitation. However, with the advent of diode lasers, some two decades later, the technique, often referred to as synchronous optical pumping, became recognized as the go-to method for OPM-based Earth-field magnetometry\,\cite{kimball2013optical}. 

Because synchronous pumping relies solely on optical fields, this approach is natural for remote magnetometry. Remote sensing of magnetic fields can be achieved by placing a vapor cell in the region to be interrogated, while pumping and readout are achieved from afar using telescopes to launch the pump and probe beams and a mirror (or retroreflector) to reflect the probe laser beam back toward the source after the interaction with the spins. A sensitive remote magnetometer that can be interrogated over several kilometers of free space is desirable in several applications, including ordnance detection, perimeter monitoring, and geophysical surveys. A magnetometer based on this approach was demonstrated experimentally by Patton et al. \cite{patton2012remotely}, which we describe in this section.

\begin{figure}[htb]
    \centering
   \includegraphics[width=0.95\textwidth]{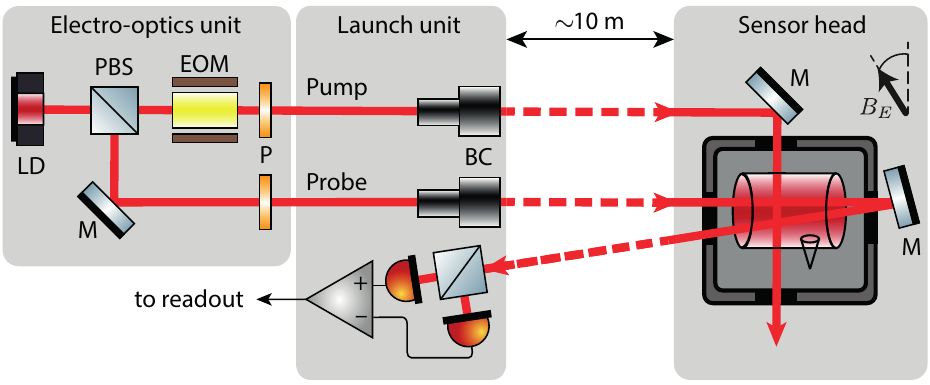}
    \caption{Schematic of the remotely interrogated magnetometer setup. LD -- laser diode, PBS -- polarizing beam splitter, EOM -- electro-optics modulator, M -- mirror, P -- polarizer, BC -- beam collimator.}
    \label{fig:remotelyInterogatedMagnetometer}
\end{figure}

The remotely interrogated magnetometer is composed of three units: an electronic unit, a launch/detection unit, and the sensor head unit, placed some 10\,m away from the other two, as shown in Fig.\,\ref{fig:remotelyInterogatedMagnetometer}. Let us briefly detail the working principle of this magnetometer by going over the function of each units. In the electro-optics unit, light emitted by a laser diode is split with a polarizing beam splitter to form a pump and a probe beam. The pump-beam is modulated at twice the Larmor frequency $\omega_L$ with the help of an electro-optic modulator, which, with a polarizer placed right after, forms intensity modulation. This allows to measure magnetic field whose vector is not orthogonal to the pump axis \cite{pustelny2006nonlinear}. The pump and probe beams are directed toward the sensor head after being collimated within the launch/detection unit. The sensor head consists of an antirelaxation coated vapor cell filled only with enriched $^{87}$Rb, heated up to 37.5$^\circ$C with AC resistive heating. The sensor is placed in the Earth magnetic field, for which the vector is here titled 32$^\circ$ away from the pump beam axis. After passing through the cell, the probe beam is reflected with a mirror placed at the back of the sensor head. The probe beam travels through the cell in a double-pass configuration before being picked back up at the launch/detection unit. 

The readout of the magnetometer is achieved with balanced polarimetry inside the launch/detection unit. Resonances appear at $1\times$ and $2\times$ the Larmor frequency when sweeping the modulation frequency of the pump-beam intensity. Indeed, modulating the pump-beam intensity at twice the Larmor frequency creates atomic alignment inside the $^{87}$Rb vapor, producing time-varying optical rotation of the probe-beam polarization \cite{yashchuk2003selective}. Here, the resonance at the Larmor frequency is due to the magnetic field being titled away from the probe-beam propagation vector \cite{pustelny2006nonlinear}.
The Larmor frequency can be related to the magnetic field with
\begin{equation}
    \omega_L = g_F\mu_B B_e \equiv \gamma B_e, \label{eq:Larmor}
\end{equation}
where $B_e$ is the Earth magnetic field magnitude, $g_F\approx \pm 2/(2I+1)$ is the Land\'e factor for $^2S_{1/2}$ hyperfine states (positive for the higher $F$ and negative for the lower), $\mu_B \approx 1.4$\,MHz/G is the Bohr magneton and $\gamma=0.7\,$MHz/G is the gyromagnetic ratio of $^{87}$Rb.\footnote{Due to the 
%coupling between the nuclear spin and electron spin, and 
nuclear Zeeman effect, the two ground hyperfine states of $^{87}$Rb, i.e., $^2S_{1/2}\, F=1$ and $^2S_{1/2}\, F=2$, have gyromagnetic ratios with slightly different magnitudes, which are $|\gamma_{F=1}|\approx 0.7024$\,MHz/G and $\gamma_{F=2}\approx 0.6996$\,MHz/G, respectively. } At Earth field, the two resonances at $\omega_L$ and $2\omega_L$ split into sets of transitions because of nonlinear Zeeman shift. This can be calculated with the Breit-Rabi formula \cite{breit1931measurement,auzinsh2010optically} which expresses the splitting of the ground-state energy levels with quantum numbers $|F=I\pm1/2,m_F\rangle$
\begin{equation}\label{eq:breit-rabi}
E(F,m_F)=-\frac{A_{\text{hfs}}}{4}-g_I \mu_B m_F B \pm \frac{A_{\text{hfs}}(I+\frac{1}{2})}{2}\sqrt{1+\frac{4m_F\chi}{2I+1}+\chi^2},
\end{equation}
where $A_{\text{hfs}}$ is the hyperfine structure constant, and $\chi$ is the perturbation parameter given by
\begin{equation}
\chi=\frac{(g_I+g_J)\mu_B B}{A_{\text{hfs}}(2I+1)}.
\end{equation}
Note that the effect of the Zeeman splitting and heading errors can be mitigated with the use of all-optical spin-locking techniques \cite{bao2022All-optical}.

\begin{figure}[htb]
    \centering
    \includegraphics[width=0.8\textwidth]{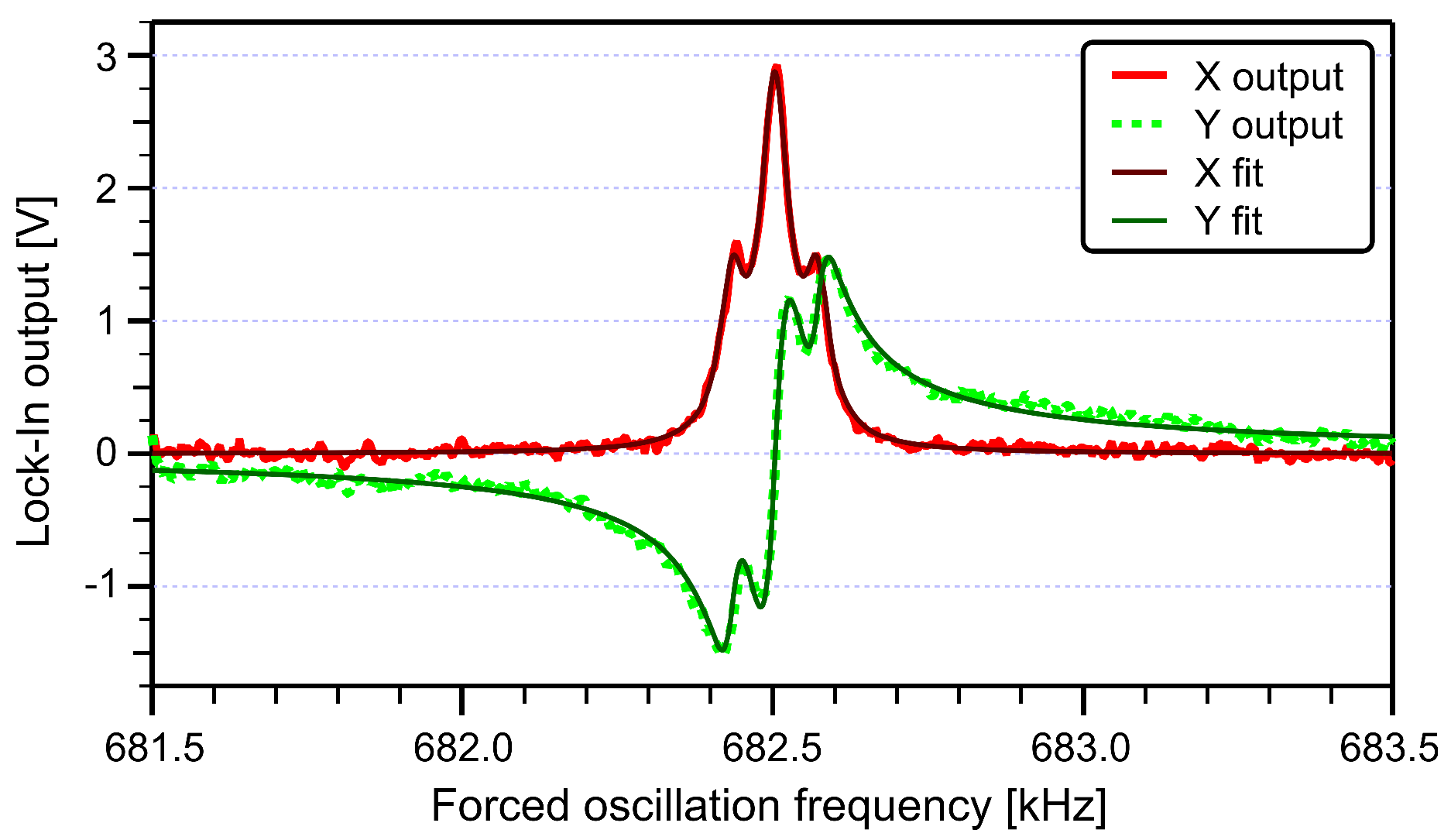}
    \caption{Resonance at $\omega_L$ recorded with the magnetometer as a function of the EOM modulation frequency. The resonance is split into a set of three resonances because of nonlinear Zeeman shift, see the text. Solid lines are fits with a three-Lorentzian magnetic-resonance spectrum predicted by the Breit-Rabi formula. Reprinted 
%    \EK{with permission} 
from Ref.\,\cite{patton2012remotely}.}
    \label{fig:remoteMagResonance}
\end{figure}

The splitting of the $\omega_L$ resonance as a function of the modulation frequency is shown in Fig.\,\ref{fig:remoteMagResonance}. Fitting of the recorded resonances with three Lorentzian functions yield a central peak at 682 504.318\,Hz with a fit uncertainty of 0.050\,Hz which suggest a field uncertainty of 3.5\,pT. Observing the sensor noise around the central peak frequency and converting the frequency to magnetic field noise with Eq.\,\eqref{eq:breit-rabi} yield a sensitivity of $5.3\,\text{pT}/\sqrt{\text{Hz}}$ in the range 1 to 50\,Hz. Recently, high-performance ($\sim 1\,\text{pT}/\sqrt{\text{Hz}}$) remote detection magnetometry was also realized with microfabricated cells \cite{LeviSPIE2023,levi2023remote}.

Replacing the mirror with a true retroreflector could enable adjustment-free remote magnetometry. We also note that, for remote magnetometry, a collinear geometry for the pump and probe is more desirable. In principle, the remotely interrogated magnetometer presented here could be operated in such way, or even with a single beam configuration \cite{shah2009Spin-exchange} as the effect of the double-pass configuration is similar to increasing the cell length, and therefore the rotation angle of the probe-beam polarization, by a factor of two \cite{Budker2002RMP}. This technique can be extended to distances above hundred meters, however noise from atmospheric seeing increases with the distance and can become a significant issue \cite{buck1967atmospheric}. Beyond this distance scale, adaptive optics techniques may become necessary to mitigate the effects of atmospheric turbulence and retain magnetometric sensitivity.

To conclude this section, we note that there exists an alternative to using a retro-reflector for the remote measurement of magnetic fields based on polarization rotation. ``Selective reflection'' at the interface between glass and a resonant vapor has been shown to carry information on ground-state Zeeman coherences \cite{Weis1992observation}. While this elegant approach seems attractive for remote sensing because the reflected beam propagates back to the source, only limited amount of work seems to have been done in this direction \cite{Klinger2020Proof}. 
 
 \subsection{Magnetometry based on the effect of coherent population trapping} \label{sec:CPT-marnetometry}

The effect of coherent population trapping (CPT) \cite{arimondo1996} in alkali vapors provides an alternative implementation of an atomic magnetometer.  Although the ultimate sensitivity with this approach is typically not as high as that of magnetometers based on direct detection of Larmor precession, as the sensitivity surpassing $\text{fT}/\sqrt{\text{Hz}}$ level was reported in a spin-exchange-free (SERF) regime  \cite{kominis} and projected for magnetometry based on nonlinear Faraday effect \cite{budker2000sensitive}, the CPT-based approach may have advantages in applications where compactness and low power consumption are important.

The CPT method relies on measurements of the frequency of narrow dark resonances. These arise when two ground-state levels are coupled to a common excited state by two coherent optical fields in so-called $\Lambda$-type level configuration and their frequency difference matches the ground-state splitting. As a result, the atoms are pumped into a coherent superposition of Zeeman sublevels, typically belonging to different hyperfine components of the ground state. This superposition is known as the coherent dark state since atoms no longer absorb and re-emit resonant light due to destructive interference of two excitation channels. The ultimate width of the coherent resonances is determined by the ground-state relaxation rate. Thus, its typical spectral width is orders of magnitude smaller than the natural linewidth of the strong optical transitions used for preparing this superposition.

\begin{figure} 
    \centering
   \includegraphics[width=\textwidth]{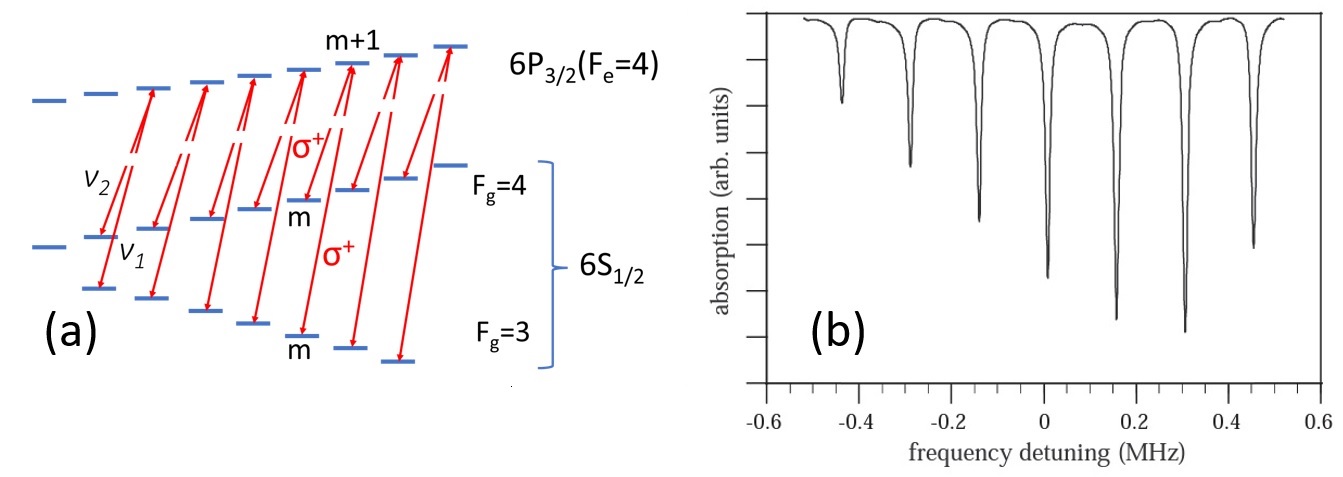}
    \caption{(a) Schematic and (b) experimentally observed Zeeman-split CPT resonances in Cs vapors in a longitudinal magnetic flux density of 21\,$\mu$T with $\sigma^+-\sigma^+$ excitation produced by two coherent optical fields with frequency offset $\delta = \nu_1-\nu_2-\Delta_{HF}$, where $\Delta_{HF}$ is the hyperfine splitting of the $6S_{1/2}$ ground state. The selection rules allow seven coherent dark resonances.  From Ref.\,\cite{nagel_EurLett_1998}. }
    \label{Fig:Zeeman_CPT}
\end{figure}

In the presence of an external magnetic field, a dark resonance splits into several components as was shown in one of the first realizations of CPT in Rb vapors  \cite{akulshin_91}. The number of components and their spectral positions depend on many parameters, among which the most important are the direction and strength of the local magnetic field. As an example, Fig.\,\ref{Fig:Zeeman_CPT} shows seven $\Lambda$-type configurations formed in Cs atoms by circularly polarized resonant radiation $\nu_1$ and $\nu_2$ that couple the Zeeman sub-levels of the two ground-state hyperfine levels and give rise to well-resolved coherent resonances of reduced absorption.

Initially, it was assumed that CPT resonances would be immune to power broadening, and therefore, a CPT-based magnetometer could provide improved sensitivity over conventional optically pumped counterparts \cite{scully-magnetometer}. Later it was found that other factors, such as frequency noise of applied laser light and  the AC Stark effect coupled with laser-light fluctuations limit the sensitivity of CPT magnetometers \cite{Matsko_PRA2000}. Nevertheless, it was shown that for both AC and DC magnetic fields, sensitivity values of the order of $1\,\text{pT}/\sqrt{\text{Hz}}$ are readily achievable \cite{nagel_EurLett_1998}.

% Attention to CPT-based magnetometry was revived with the advent of a new type of diode lasers. The outstanding properties of vertical cavity surface-emitting lasers (VCSELs), which are characterized by  the spectral purity of their emission, large modulation bandwidth,  threshold currents below 1\,mA, high efficiency of converting electrical energy into photon flux, excellent reliability and low cost, make them ideal optical sources for data processing and sensing \cite{vcsel}. These unique lasers, combined with microfabricated alkali vapor cells, have enabled the development of chip-scale atomic clocks based on the CPT effect in Rb or Cs vapors in response to a strong demand from the communications industry for compact and low-power frequency references.

%\AAk{If necessary, I can add a paragraph about the specific benefits of VCSELs to CPT preparation.}
Attention to CPT-based magnetometry was revived with the advent of a new type of diode lasers. The outstanding properties of vertical cavity surface-emitting lasers (VCSELs) make them ideal sources of coherent radiation for remote sensing \cite{vcsel}. The VCSELs are a type of semiconductor lasers that emit coherent radiation from the top surface of the chip, in contrast to the conventional edge-emitting diode lasers. The active region, typically sandwiched between two semiconductor Bragg mirrors, has a total thickness of only a few micrometres. This ensures low threshold current, below 1 mA, and reduced power consumption. The VCSELs with an emitting area with typical dimensions of few microns  produce high-quality and symmetric beam profiles that simplify coupling of the output radiation to optical fibers. In addition, VCSELs can be efficiently modulated at high radio frequencies generating mutually coherent optical fields, which is important for CPT-based applications. Although the output power of multimode VCSELs can reach tens of watts, the output power of VCSELs generating radiation in the fundamental transverse mode is in the range of 0.5–5 mW, which is quite sufficient for creating CPT in alkali vapours. Other important features of VCSELs are excellent reliability and long lifetime due to substantially lower optical intensities at the output facet, compared with edge-emitting diode lasers. Therefore, they can be operated at higher temperatures, which would normally increase the risk of catastrophic facet failure. The most common emission wavelengths of VCSELs are in the range of 750 – 980\,nm that perfectly fits with the strong absorption lines in alkali atoms used for optical magnetometry. These unique lasers, combined with microfabricated alkali vapor cells, have enabled the development of chip-scale atomic clocks based on the CPT effect in Rb and Cs vapors in response to a strong demand from the communications industry for compact and low-power frequency references.

Owing to the common underlying physics of the CPT effect, the technological advances in the development of chip-scale atomic clocks are easily transferred to CPT-based magnetometers. The most significant difference between the two types of devices is that the coherent dark superposition for magnetometers is excited among Zeeman sublevels that are sensitive to magnetic field, while $m_F$ = 0 sublevels are often used for atomic clocks. %\DB{Sasha, please have a look at push-pull.}

\begin{figure}
    \centering
   \includegraphics[width=\textwidth]{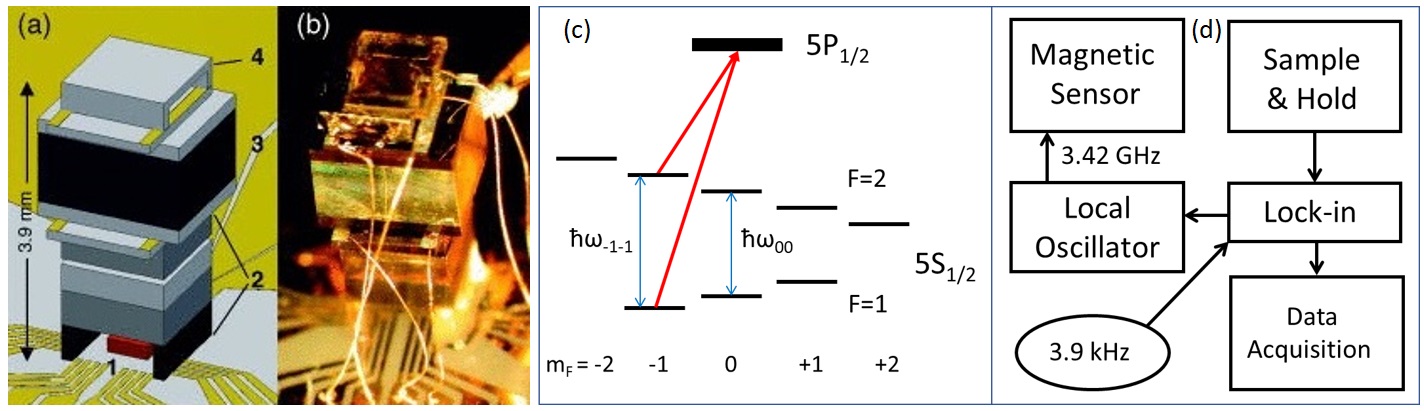}
    \caption{(a) Schematic of the chip-scale magnetic sensor. 1 - VCSEL, 2 - optics package including neutral-density filters, a microlens, a $\lambda$/4 waveplate, and a neutral-density filters, 3 - $^{87}$Rb vapor cell with transparent heaters above and below it, and 4 - (a) photodiode assembly. (b) Photograph of the magnetic sensor. (c) Energy-level diagram of  $^{87}$Rb indicating the CPT transition scheme. (d) Experimental setup for detecting the magnetic flux density. The dashed arrow indicates that the lock-in amplifier can provide feedback to the local oscillator. From Ref.\,\cite{chip-scale}. }
    \label{Fig:Chip-scale}
\end{figure}

The first low-power chip-scale magnetometer based on the CPT resonance in alkali vapors  was realized at the USA National Institute of Standards and Technology (NIST) \cite{chip-scale}. The schematic of the chip-scale magnetic sensor, energy-level diagram of $^{87}$Rb with the CPT transitions, and the scheme for detecting the magnetic flux density are shown in  Fig.\,\ref{Fig:Chip-scale}. The two mutually coherent optical fields required for establishing the CPT were generated by modulating the pump current, which greatly simplifies the optical scheme and drastically reduces the power consumption. The magnetic sensor had a volume of 12\,mm$^{3}$ and dissipated 195\,mW of power. It was possible to detect the magnetic flux density with a sensitivity of $50\,\text{pT}/\sqrt{\text{Hz}}$ at 10\,Hz.

%\DB{The discussion below needs to be updated with a more recent review by John Kitching, see \cite{kitching2018chip-scale}.}

Generally, a relatively small fraction of the alkali atoms can be pumped into a dark superposition state. It was suggested to use a so-called ``push-pull’’ pumping method to significantly improve the performance of CPT-based atomic clocks \cite{happer_2004}. This technique makes it possible to transfer almost all atoms into the $m=0/m'=0$ superposition state, because its resonance frequency is insensitive to magnetic fields. This method uses a sequence of pulses of D1-resonant light with left and right circular polarization, which alternate with the Larmor frequency. Later, push-pull pumping was applied to magnetically sensitive resonances \cite{weis_2014} as well. The magnetization of spin-oriented alkali atoms prepared by optical pumping with circularly polarized laser light precesses around a static magnetic field.  After half the Larmor period, the spin polarization changes sign, and so does the light polarization, further reinforcing spin polarization.
This process repeats periodically until a steady-state precessing polarization of constant amplitude is reached (see also Sec.\,\ref{subsec:LGS_experiment}).
%Thus, switching the polarization of laser light at the Larmor frequency effectively preserves the polarization of the atomic sample. 
An atomic magnetometer based on the push-pull pumping with a room temperature paraffin-coated Cs vapor cell with ultimate
shot-noise-limited sensitivity below  $20\,\text{fT}/\sqrt{\text{Hz}}$ was reported in  \cite{weis_2014}. An interesting feature of the push-pull approach is that by using magnetically sensitive substates it enables hybrid devices, for instance, an atomic clock combined with a magnetometer.  

% \RZ{We should find a place to mention this work on sodium magnetometry with synchronous modulation  of two photons (CPT) resonance \cite{grewal2020APL}: "We have demonstrated a technique for remote magnetometry using fluorescence
% measurements in a sodium cell. This technique utilizes two-photon resonance instead of
% single-photon resonance to generate magnetic resonances with synchronous modulation of …".}

The following years saw further improvements in miniaturization, power dissipation, and magnetic field sensitivity. The CPT-based magnetometers may be a good choice when power consumption and volume matter. The concept of miniaturization has been extended also to other types of magnetometers. There is an obvious trade-off between the size and power consumption of a magnetometer and its characteristics.   The design, fabrication, and performance of miniature devices, as well as the prospects for their possible applications, are discussed in a recent comprehensive review \cite{kitching2018chip-scale}.

There is enormous potential for remote sensing of magnetic fields over long distances using disposable battery-powered millimeter-sized devices. As an example, CPT magnetometers are used on board the Jupiter Icy Moons Explorer mission of the European Space Agency \cite{Amtmann2024}. According to the results of careful calibrations conducted in the geomagnetic Conrad Observatory, the magnetometers are able to measure magnetic field strengths down to 100\,nT with a systematic error less than 0.2\,nT \cite{Ellmeier2024}.

%The intrinsically scalar CPT magnetometer can be turned into a true vector magnetometer by adding a set of modulated bias fields---a standard approach in magnetometry. Chip-scale CPT-based magnetometers can also operate near zero magnetic field \cite{Hong_Chip-Scale}.

%\DB{Shall w stop here and just add a few concluding sentences of CPT?} \AAk{Yes!!}

In conclusion of this section, we note that the physical mechanisms underlying CPT-based magnetometry are similar to those of polarization-modulation-based LGS magnetometry discussed in Sec.\,\ref{subsec:LGS_experiment}. Remote magnetometry can be achieved with low-power CPT magnetometers described in this section by placing a complete sensor where the field needs to be measured using, for example, UAVs or UUVs. With detection based on fluorescence of alkali atoms, only a part of the sensor, the vapor cell, need to be placed in the measurement region while the launch of the laser beams and detection is carried at user's location. This later approach was demonstrated in a proof-of-principle tabletop system with a  sensitivity of %Grewal et al \cite{grewal2020APL} who reported a sensitivity of 
45\,pT$/\sqrt{\text{Hz}}$ \cite{grewal2020APL}. 

Other tabletop magnetometry experiments were performed with detection based on fluorescence \cite{fan2018magnetometry,grewal2020magneto} or absorption \cite{ding2022magnetometry}  of sodium atoms. The results of these studies could be useful for informing on-sky measurements with mesospheric sodium, discussed in the next section.

\section{Laser guide star magnetometry based on mesospheric atoms} \label{sec:LGS-mag}

 % ------------------------------ Log -----------------------------%
%
%Log:
%\begin{itemize}
%    \item \FP{April 13: Added properties sodium layer and applications}
%    \item \FP{April 15: Added atmospheric profiles. Investigated new realization of the experiment: no new experiments after Fan (2019). Added up-to-date table of experiments and relevant parameters (must double check the numbers). Retrieved science information on thermosphere proceses from slides of Jürgen Matzka (must be acknowledged, also ask if there are updates from the experiment in Norway)}
%    \item \FP{June 21: Summer! Added information under section  ''Concept".}
%    \item \FP{July 1st: Went over section 4.1 and first paragraph 4.2.}
%\end{itemize}
 % ----------------------------End Log -----------------------------%

%\EK{check "13.3 Magnetometry with mesospheric sodium" in \cite{kimball2013optical} to not miss any early material on Mesospheric sodium magnetometry}

\subsection{The concept}

A laser guide star (LGS) is an artificial source of light created in the upper mesosphere between 85~km and 100~km altitude, conceived as a beacon to guide an adaptive optics (AO) system for compensating atmospheric-induced aberrations in ground-based astronomical observations~\cite{happer1994atmospheric,Fugate:1991b}.\footnote{Although first proposed in the scientific literature by French astronomers Foy and Labeyrie in 1985~\cite{Foy:1985}, laser guide stars were developed by the US Department of Defense under a classified adaptive-optics program since 1981, and their first experimental results were published only by 1991~\cite{Humphreys:1991, Duffner:2009}.} An LGS is created by resonant optical excitation of neutral sodium atoms in the mesosphere using a laser beam launched from a ground-based telescope. The laser is typically tuned to the D$_2$ absorption line with a wavelength of 589.1~nm for maximum efficiency. Fluorescence from the LGS is collected with another telescope, often used for astronomical observations, and then fed into the AO system enabling real-time correction of the incoming light wavefront. The performance of the AO system, i.e. the capability to deliver near diffraction-limited images to a science instrument, improves with a brighter LGS. In order to maximize the photon return flux from LGS, significant research and development has been done for the last 30 years, exploring the physics of light interaction in the mesosphere~\cite{Miloni:1999,Holzloehner2010AA,Rochester:2012} and pushing innovation in laser technology~\cite{Denman:2005,Feng:2009}. Furthermore, advanced concepts such as polychromatic laser guide stars for compensation of atmospheric tilt~\cite{Foy:1995}, and laser photometric ratio star to be used as precise spectrophotometric calibrator~\cite{albert2021preciseI,albert2021preciseII} were proposed.

\begin{figure}[t]
\begin{center}  
   \includegraphics[width=0.6\linewidth]{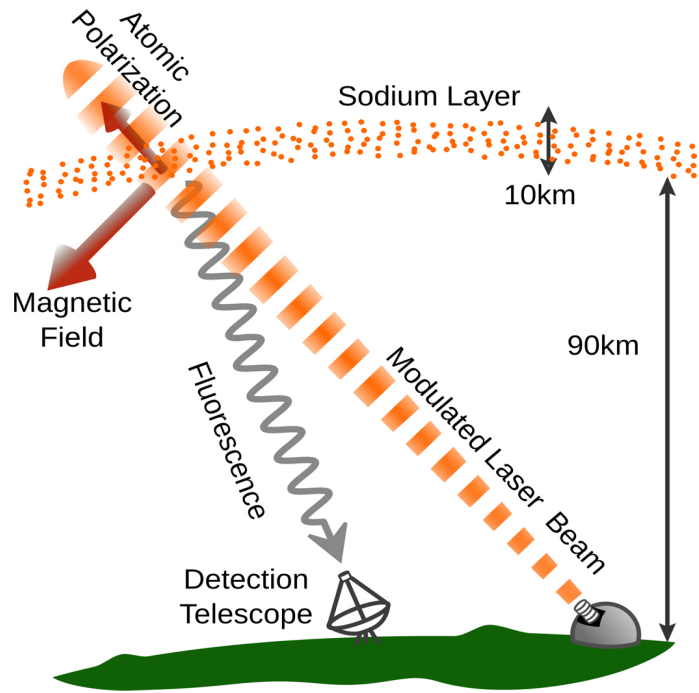}
    \caption{Concept of laser guide star magnetometry. A modulated laser beam optically pump mesospheric sodium in a region nearly perpendicular to the local magnetic field. Modulating around the Larmor frequency gives rise to a resonant feature in fluorescence which is detected with a ground-based telescope. The strength of the magnetic field is proportional to the measured Larmor frequency. Image taken from \cite{Higbie2011}.}  \label{fig:LGSconcept}
\end{center}
\end{figure}

It was found that the brightness of LGS optically pumped with continuous wave lasers, depends on the pointing direction of the laser beam in the sky, mainly as a result of precession of the atomic polarization in the geomagnetic field lines~\cite{Moussaoui2009}. The redistribution of atomic populations over magnetic sublevels can result in reduced photon-absorption probability, leading to a fainter LGS. Although this effect is detrimental for AO, it can be exploited to measure the surrounding magnetic field as originally proposed by Higbie et al~\cite{Higbie2011} in 2011. The principle of LGS magnetometry relies on the measurement of the precession frequency of atomic polarization, proportional to the strength of the magnetic field, as first demonstrated by Bell and Bloom~\cite{bell1961optically} in 1961 employing Cs and Rb gas cells. In LGS magnetometry, free sodium atoms are optical pumped using a circularly-polarized laser beam pointing at an angle ideally perpendicular to the local magnetic field in the mesosphere, causing polarization of the ground state\footnote{The process of pumping atoms synchronously with Larmor precession is referred to it as synchronous optical pumping or optically driven spin precession.} and precession around the field lines at the Larmor frequency. The rate of optical pumping must be modulated (either in intensity, polarization, or optical frequency) near the Larmor frequency, increasing the degree of atomic polarization, in turn, modifying the fluorescence which is observed with a receiver telescope on the ground. Scanning the modulation frequency results in a resonant signal; a sharp increase if the laser wavelength is set to the D$_2$ line, or a decrease if set to the D$_1$ line. The modulation frequency at which the resonant peak (dip) occurs can be identified and measured accurately, from which the strength of the magnetic field is derived following Eq.\,\eqref{eq:Larmor} (Section~\ref{sec:LightReflectionMag}), using the gyromagnetic ratio of the $\ket{F=2}$ sodium ground state $\gamma=699.812$~kHz/G. A schematic of an implementation of this technique is shown in Fig.\,\ref{fig:LGSconcept}. 

\subsection{The geomagnetic field}

The magnetic field of the Earth (the geomagnetic field) originates from different sources having a wide range of spatial scales, varying at different time scales. The geomagnetic field is used as a guide for navigation by imparting a directional preference to the needle of a compass. Studying the geomagnetic field has contributed to the formulation of the plate-tectonics theory \cite{Runcorn:1957}, the discovery and mapping of underground structures \cite{Theocaris:1996}, and better understanding of the Earth's space environment \cite{Paschmann:2005}.

The generally accepted theory of the origin of the geomagnetic field is that it is generated in the fluid outer core of the Earth by a self-exciting-dynamo process. Convective motion of the conducting iron surrounding the inner core rotates due to the Coriolis force, and as a result of the rotating conductive flow, a magnetic field is induced~\cite{McNish1940,Buffett:2007}. The core magnetic field extends outside the solid Earth up to the magnetopause at a distance of about 10 times the Earth radius; a boundary where the pressure from solar wind (the constant flow of charged particles streaming off the sun) and the geomagnetic field are balanced. The magnetosphere shields our planet from solar and cosmic particle radiation, preventing the erosion of the atmosphere by the solar wind. For these reasons, the geomagnetic field and the magnetosphere play an essential role in the development of life on Earth~\cite{erdmann2021}.

The orientation of the geomagnetic field vector at any point is defined by its inclination $I$ and declination $D$ angles, where inclination angle is measured from the horizontal plane to the field vector (positive downwards), and declination angle is measured clockwise from true north to the horizontal component of the field vector (seen from above). The size of the field vector is the total intensity $F$. Globally, $D$ can take values between $\pm180^\circ$ and $I$ can take any value between $\pm90^\circ$, while $F$ ranges between $20$ and $67\,\mu$T~\cite{WMM:2020}.

%\begin{figure}[h]
%\begin{center}  
%   \includegraphics[width=0.6\linewidth]{LGS/PlaceHolder_CoordinateSystem.png}
%    \caption{Reference system of the geomagnetic field.}  \label{fig:BfieldReference}
%\end{center}
%\end{figure}

\subsection{Geomagnetic models}

%There are mainly two global geomagnetic models. 
One of the two leading models, the International Geomagnetic Reference Field (IGRF) model was introduced by the International Association of Geomagnetism and Aeronomy (IAGA) in 1968 in response to the demand for a standard spherical harmonic representation of the Earth's main field. The model is updated at 5-year intervals, the latest being the 13th generation (IGRF13)~\cite{IGRF13}. The IGRF incorporates several mathematical models of the Earth's main field and its annual rate of change (secular variation) and it represents only the main (core) field without the influence of external sources. The U.S. National Oceanic and Atmospheric Administration (NOAA) introduced the World Magnetic Model (WMM); it is the standard model for the UK Ministry of Defence and the US Department of Defense. The latest version was released in 2020~\cite{WMM:2020}. The WMM is a model of the core and large-scale crustal fields only. 

In both models, the main magnetic field $\mathbf{B}_\text{m}$ is a potential field and therefore can be written in geocentric spherical coordinates (longitude $\lambda$, latitude $\phi$, and radius $r$) as the negative spatial gradient of a scalar potential
\begin{equation}
\mathbf{B}_\text{m}(\lambda,\phi,r,t) = - \nabla V(\lambda,\phi,r,t)\,. \label{eq:ch4:scalar_potential}
\end{equation}
The scalar potential can be expanded in terms of spherical harmonics:
\begin{equation}
\nabla V(\lambda,\phi,r,t) = a \sum_{n=1}^N \sum_{m=0}^n \left( \frac{a}{r} \right)^{n+1} \left[ g_ n^m (t) \cos(m \lambda) + h_n^m (t)\sin(m \lambda) \right] \breve{P}_n^m (\sin \phi)\,, \label{eq:ch4:WMM2015_model}
\end{equation}
where $N=12$ is the degree of expansion of the WMM and IGRF models, $a$ is the geomagnetic Earth's reference radius ($6\,371\,200$~m), and $g_ n^m (t)$ and $h_n^m (t)$ are the time-dependent Gauss coefficients of degree $n$ and order $m$ describing the Earth's main magnetic field. $\breve{P}_n^m (\mu)$ are the Schmidt semi-normalized Legendre functions associated with the real number $\mu$. The WMM2020 and IGRF13 models provide two sets of Gauss coefficients for the calculation of the main field and secular variations, respectively. In order to solve Eq.\,\eqref{eq:ch4:scalar_potential} detailed procedures are described in the documentation of each model.

The accuracy of the estimated total intensity is tied to changes in the fluid flow in the Earth's outter core, leading to unpredictable changes in the geomagnetic field. The models account for long-spatial scales of the internal Earth magnetic field. This means that variations in the geomagnetic field generated from the crust and upper mantle, as well as from ionospheric origin, are under represented in the models. In practice, significant differences between the values predicted by the models and measurements at a given point on Earth may occur. For example, daily variations of the magnetic field lies typically on the range of tens of nanotesla, and they are caused by electric currents flowing on the sunlit side of the ionosphere ~\cite{Yamazaki:2017}. Local anomalies on the order of hundreds of nanotesla arising from geologic features, are also not characterized in the model. The accuracy in the total field $F$ provided by WMM2020 model is 145 nT. Because none of these models include local magnetic disturbances, it is important to sample and monitor the geomagnetic field at small-spatial and short-time scales. Continuous measurements of the geomagnetic field by a global network of geomagnetic observatories~\cite{Love:2013} contributes to regularly update the aforementioned geomagnetic models in periods of five years.

%\textbf{Ionospheric electric currents}\\
%horizontal current  sheet at 120 km altitude (approx. boundary between atmosphere and space)
%currents produce magnetic field that is measured at ground and satellites\\
%\textbf{Auroral zone electric currents}\\
%small scale structure, approx. 40 km  (comparable to auroral structure)\\
%only detectable by magnetic field sensor within 20 km (half wavelength) distance\\
%but: basically no magnetic data from 30 to 300 km altitude available\\
%\textbf{Principles mesospheric magnetometer}\\
%measures magnetic field strength\\
%ground based, measures remotely in the natural sodium layer at 90 km\\
%potential to detect ionospheric current structures down to 40 km scale size

\subsection{Properties of the sodium layer}

The ``sodium layer" (in which sodium is still a trace-amount minority species) is composed of a mixture of atomic sodium with other atomic and molecular species between 82 km and 105 km altitude, at the interface of the mesosphere and the thermosphere atmospheric shells. Ablation of meteorites in the thermosphere together with chemical reactions provides a constant source of Na and other metals such as K, Fe, Mg, Ca, and Si. Between 82\,km and 105\,km, these metals exist as a layer of neutral atoms. Above this layer, metals are ionized, while below 82\,km they form oxides, hydroxides, and carbonate compounds. Nevertheless, the two most abundant species are molecular oxygen (O$_2$) and nitrogen (N$_2$) that surpass by several orders of magnitude the density of sodium and other metals. The properties of these atomic species in the mesosphere were discussed in \cite{happer1994atmospheric} and more recently in \cite{Yang2021atomic}, and are summarized in Table\,\ref{Tab:mesosphere_species}. The steep increase of oxygen density towards lower altitudes makes the recombination process faster, which defines the sharp boundary of the sodium layer at altitudes of about 82\,km. It takes about four years for sodium compounds to finally reach the Earth surface \cite{plane2015mesosphere}.

\begin{figure}[t]
\begin{center}  
   \includegraphics[width=0.6\linewidth]{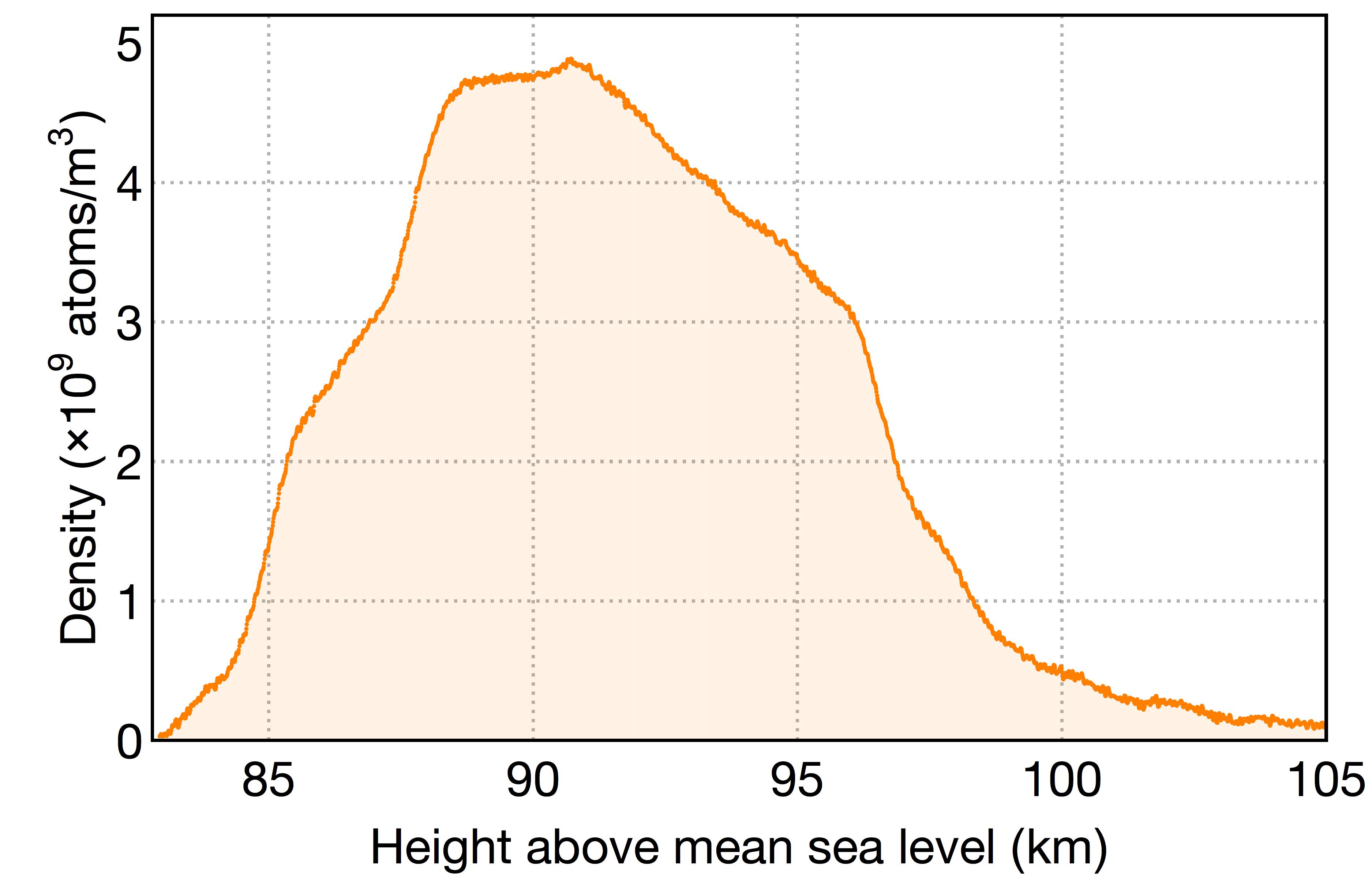}
    \caption{Reference vertical sodium density profile obtained with the Large Zenith Telescope LIDAR facility in Vancouver, (BC, Canada), normalized with a sodium column abundance of $C_\text{Na}=5.0 \times 10^{13}$~atoms$/\text{m}^2$. Data courtesy of J. Hellemeier and P. Hickson, University of British Columbia.}  \label{fig:ch2:Na_profile}
\end{center}
\end{figure}

The sodium layer has been actively studied since the 1970s using lidar (light detection and ranging) in order to characterize the sodium properties and variability \cite{Gibson:1979,Megie:1978}, to measure mesospheric temperature and winds \cite{Dunker:2015,She:1994}, and ultimately for studying large-scale atmospheric couplings and gravity waves \cite{Xu:2004}. Motivated by the implications on adaptive optics, the sodium layer has been characterized at different sites over the globe in order to determine its properties and the time-variability on short and seasonal timescales \cite{Michaille2001,Neichel:2013,Pfrommer:2014,Pfrommer:2010}. From a statistical analysis of 35 years of data from lidar observations of the sodium layer in Sao Paulo, Brazil (latitude 23\degree\ South), the average centroid of the sodium layer is 92.09~km, with a thickness of 11.37~km and a column density of $4-5 \times 10^{13}$~atoms/m$^{2}$ \cite{Moussaoui:2010, Yang2021atomic}. In fact, the average global mass of sodium in this layer can be estimated to be on the order of 1 ton. The properties of the sodium layer (abundance and morphology) are, however, highly variable. To date, the highest temporal- and spatial-resolution characterization of the sodium layer was carried out by Pfrommer and Hickson \cite{PfrommerThesis:2010} using a sodium lidar with a 6\,m liquid mirror receiver telescope at the Large Zenith Telescope (LZT) in British Columbia, Canada. Their work showed, for example, sodium centroid vertical variation of $\sim$23\,m/s RMS (root-mean-square) on a 1\,s timescale, the presence of multiple sodium layers with lifetimes of several hours, sporadic sodium layers (SSL) that can exceed the mean sodium abundance by one order of magnitude, the identification of Kelvin-Helmholtz rolls and coherent short-period gravity waves in the mesosphere, and the classification of seven typical sodium-density profiles. Their studies on the vertical variation of the sodium layer centroid were particularly important to assess its impact on the wavefront error in future extremely large telescopes \cite{Herriot:2006}. A typical sodium density profile obtained with the LZT is shown in Fig.\,\ref{fig:ch2:Na_profile}.

\begin{table}[htp]
\begin{tabular}{ccccc}
\hline\hline
Species & \makecell{Column density\\ (atoms/m$^{2}$)} & \makecell{Primary\\ E1 transition} & \makecell{Ground state $\rightarrow$ \\Exited state} & Reference\\
\hline
O &$6.5 \times 10^{21}$ & $2\times 226$\,nm & $\textrm{(2p)}^3P  \rightarrow \textrm{(3p)}^3P$& \cite{Yang2021atomic,swenson2018vertical,dogariu2018high} \\
H & $2.3 \times 10^{18}$& $2\times 205$\,nm &$\textrm{(1s)}^2S  \rightarrow \textrm{(3s)}^2S$ & \cite{Yang2021atomic,dogariu2018high} \\
N & $1 \times 10^{18}$& $2\times 211$\,nm & $\textrm{(2p)}{^3}S_0  \rightarrow \textrm{(3p)}^4D_0$& \cite{Yang2021atomic,adams1998two}  \\
Fe & $10.2 \times 10^{13}$& 372\,nm& $a^5D_{4}  \rightarrow z^5F_{5}$ &\cite{Yang2021atomic,gardner2004performance,gelbwachs1994iron} \\
Na & $4.0 \times 10^{13}$ & 589\,nm & $3S_{1/2}  \rightarrow 3P_{3/2}$ & \cite{Yang2021atomic,plane2015mesosphere} \\
Mg & $1.0 \times 10^{13}$& 285\,nm & $3^{1}S_{0}  \rightarrow 3^{1}P_{1}$&\cite{Yang2021atomic,whalley2011kinetic, plane2010magnesium} \\
Ca & $3.4 \times 10^{11}$& 423\,nm &$4^{1}S_{0}  \rightarrow 4^{1}P_{1}$ & \cite{verner1996atomic, plane2015mesosphere,plane2010magnesium}\\
K & $4.5 \times 10^{7} $ & 770\,nm &$4S_{1/2}  \rightarrow 4P_{1/2}$ &  \cite{Yang2021atomic, gardner2004performance, plane2015mesosphere}\\
%\TD{\ch{Si+}} &  & & & \\
Li &$\sim 10^{6}$ &671\,nm &$2S_{1/2}  \rightarrow 2P_{1/2}$ & \cite{jegou1980lidar, vaughan2014observations} \\
\hline\hline
\end{tabular}
\caption{Parameters of the most abundant atomic species in the upper atmosphere between 82\,km and 105\,km altitude. Rubidium and cesium are not observed in the Earth atmosphere. Sodium is chosen as the species for most laser guide stars and mesospheric magnetometry because of its relatively high abundance, strong optical transitions at a wavelength with high atmospheric transmission, accessible to high-power lasers.}
%\DB{Let us please add Li, K, Rb, Cs)} \TD{Si atoms are readily oxidized such that Si species exist in \ch{Si+} and \ch{SiO}}\DB{This is a table with atomic species so an ion does not belong. I suggest discussing it in the main text if we want to or making another table for ions....} }
\label{Tab:mesosphere_species}
\end{table}

\begin{figure}[htb]
\begin{center}    
    \includegraphics[width=0.99\linewidth]{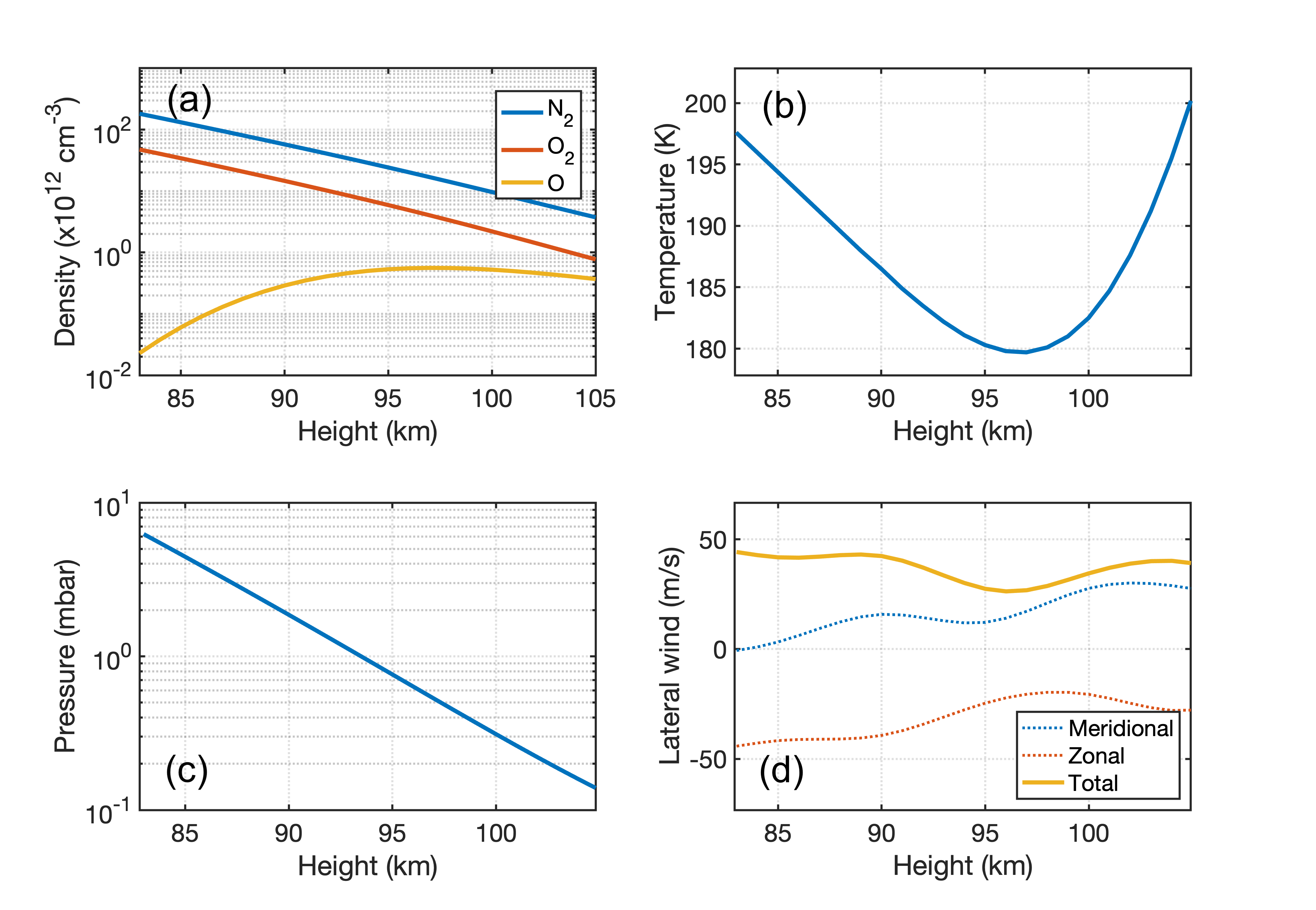}
    \caption{Example of vertical atmospheric profiles between 83\,km and 105\,km altitude above sea level for La Palma, Canary Islands (Lat. $+28.75$\degree\ Lon. $-17.90$\degree) of (a) number density of N$_2$, O$_2$ and O, (b) temperature, (c) total pressure, and (d) horizontal winds. (a -- c) Obtained from the NRLMSISE-00 model\,\cite{Picone:2002}, and (d) obtained from the HWM2014 model\,\cite{Drob:2015}.} \label{fig:ProfilesAtmosphere}
\end{center}
\end{figure}

%To do:
%\begin{itemize}
%    \item Add table of species with: abundance, ground state, excited state, lowest E1 accessible (wavelength), source of species. 
%\end{itemize}

\subsection{Sensitivity}

The spin-projection noise-limited sensitivity (atomic shot-noise sensitivity) $\delta B_\text{snl}$ of a magnetometer based on polarized atoms is determined by~\cite{budker2007optical}
\begin{equation}
\delta B_\text{snl} \approx \dfrac{1}{\gamma}\sqrt{\dfrac{\Gamma_\text{rel}}{N \tau}}\,, \label{eq:ch4:Bsn}
\end{equation} 
where $\Gamma_\text{rel}$ is the spin-relaxation rate, $N$ is the total number of polarized atoms, and $\tau$ is the measurement time. Considering a uniform laser beam profile of an area of 
$0.2$\,m$^2$ illuminating the sodium layer at a zenith angle\footnote{Zenith angle is the angle between the point directly above the observer and the pointing direction.} $\zeta=30$\degree, and a total Na column density of $\sigma_\text{Na}=4\times 10^{13}$~atoms/m$^{2}$ (at zenith), the total number of pumped atoms in the mesosphere is $N=9.1\times 10^{12}$ atoms. Assuming $\Gamma_\text{rel}=450$~$\mu$s and 1~s integration time, the spin-projection noise is $\delta B_\text{snl} = 2.2\ \text{fT}/\sqrt{\text{Hz}}$. 

However, the total number of photons that can be detected scales with the solid angle $\Omega_\text{tel}$ subtended by a receiver telescope with collecting area $A_\text{tel}$ at a distance $L$ from the sodium layer, such as
\begin{equation}
\Omega_\text{tel} = \dfrac{A_\text{tel}}{L^2}\,,
\end{equation}
with $L=(H_\text{Na} - H_\text{tel})/\cos(\zeta)$, where $H_\text{Na}$ is the vertical height of the Na centroid and $H_\text{tel}$ is the vertical height of the telescope location. For example, for $H_\text{Na}=92$~km, $H_\text{tel}=2.4$\,km (typical elevation of astronomical sites) and $\zeta=30$\degree, we have $L=103.4$\,km.

The fraction of photons collected by the receiver telescope compared to the total number of photons radiated in all directions can be approximated as
\begin{equation}
\xi = \dfrac{\Omega_\text{tel}}{\Omega_\text{iso}} = \dfrac{A_\text{tel}}{\frac{3}{2} 4 \pi L^2}\,.
\end{equation}
The factor $3/2$ in the denominator accounts for the radiation pattern of atoms pumped with circularly polarized light with respect to perfect isotropic radiation. With $L=103.4$~km and for a telescope of collecting area $A_\text{tel}=0.1$~m$^2$ the fraction $\xi=5.0\times 10^{-13}$. With this setup the fundamental atomic shot-noise sensitivity limit for mesospheric magnetometry experiments is then given by Eq.~\eqref{eq:ch4:Bsn} using $N\rightarrow N \xi$, which leads to  $\delta B_\text{snl} = 3.2$~nT/$\sqrt{\text{Hz}}$.

\subsection{Experimental realization}\label{subsec:LGS_experiment}

Successful demonstrations of mesospheric magnetometry were reported in Refs.\,\cite{Kane2018,Pedreros2018,Pedreros2018polarization,Fan2019remote} (see Table\,\ref{table:MagExperiments}) and the developments are ongoing with the goal of investigating spatial variations of the magnetic field in the mesosphere, exploring new formats of magnetometry to extend the sky coverage, and, apart from magnetometry, to evaluating the benefits of optical pumping at the Larmor frequency to boost the brightness of LGS. So far, a boost of up to 20\% in the LGS return when using modulated light was demonstrated, in agreement with theoretical simulations~\cite{bustos2018simulations}.

The first demonstration took place at the Steward Observatory (Tucson, Arizona) using the 1.55\,m  Kuiper Telescope as a receiver\,\cite{Kane2018}. A laser based on sum-frequency-generation technology delivered a 3.8\,W single-mode continuous wave beam at 589\,nm on-sky~\cite{Denman:2003}. The laser beam was circularly polarized and amplitude modulation was the chosen approach to synchronously pump mesospheric sodium. Due to the relatively high repetition rate of modulation ($\approx$ 300\,kHz) and the requirement of scanning around the Larmor frequency to characterize the magnetic resonance, using an off-the-shelf acousto-optic modulator offered a simpler solution compared to intracavity laser-pulsing techniques. The choice of the duty cycle of modulation is a trade-off between the resonance contrast (better at small duty cycle) and signal strength from the mesosphere (higher with large duty cycle). Using the LGS-Bloch extension of the AtomicDensityMatrix modelling software \cite{LGSBloch}, a compromise was found near 35\%. Despite the relatively large receiver telescope, the expected level of the signal is fairly low ($\approx 300$ counts/ms) for which an amplified photon-counting detector was used. The principle of the measurement is that the modulation rate is scanned around the expected Larmor frequency, taking as a reference the ground-based data from a nearby geomagnetic observatory, and the signal from the laser guide star is recorded on the receiver side. One important feature of this experiment was the dithering method used to suppress scintillation noise due to the effect of atmospheric turbulence in the line-of-sight propagation path\footnote{Scintillation is what sky watchers witness as ``star twinkling.''}. In this method,  the repetition rate between two modulation values was dithered by $\pm$15\,kHz at a rate of 314.159\,Hz. The dither excursion of $\pm$\,15\,kHz must be larger than the expected resonance width to obtain maximum contrast in the detected data. The dither rate must be above the cut-off frequency of atmospheric scintillation, which depends on the telescope receiver aperture, wavelength, size of the LGS, and the turbulence conditions \cite{Dravins1997a,Dravins1997b,Dravins1998}. The cut-off frequency is typically a few hundred hertz. 

The resonance is encoded in the dither-modulated signal and must be detected using lock-in detection. The Larmor frequency was found at 318.0\,kHz, corresponding to a magnetic field of 45441\,nT. The nearby United States Geological Survey (USGS) Tucson Magnetic Observatory reported a magnetic field of 47404\,nT at the time of the measurements. The difference of 4\% is consistent with a radial cubic scaling law of a dipole with an origin at the center of the Earth. The sensitivity of the magnetometer was estimated after spectral analysis of a detection signal obtained with the repetition rate fixed at the point of steepest slope of the magnetic resonance, leading to an average of 162\,nT/$\sqrt{\text{Hz}}$. 

\begin{figure}[htb]
    \centering
    \includegraphics[width=0.5\textwidth]{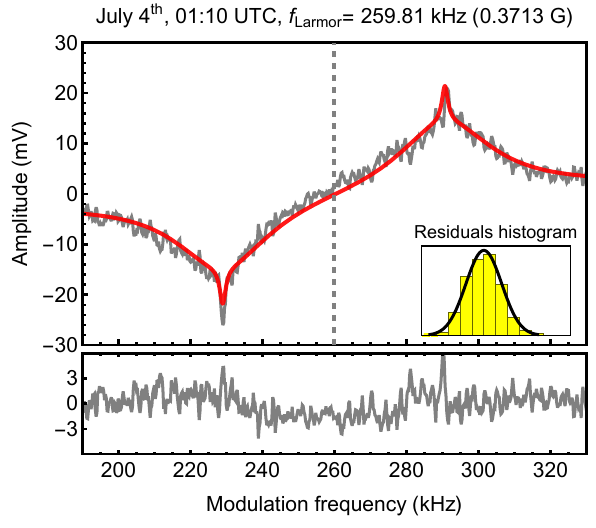}
    \caption{Magneto-optical resonance recorded in an on-sky laser guide star magnetometry experiment obtained by sweeping the frequency of the intensity-modulated laser beam (modulation duty cycle 20\%). The Larmor frequency lies in the center
between the opposite-sign peaks, which are separated by twice the dither excursion ($\delta f \approx 31\,$kHz).  The estimated laser irradiance in the mesosphere was $I^{\text{meso}}_{\text{avg}} = 17\, \text{W m}^{-2}$. A double
Lorentzian fit shows a broad and a narrow width of $\approx$\,30\,kHz and and $\approx$\,2\,kHz, respectively, consistent with two relaxation mechanisms due to velocity-changing collisions (fast) and spin-exchange collisions (slow) of sodium with N$_2$ and O$_2$ molecules. The residuals of the fits shown below obey a normal distribution according to the Gaussian fit of the residuals (histogram). Adapted from Ref.\,\cite{Pedreros2018}.}
    \label{fig:Pedreros2018Resonance}
\end{figure}

The second demonstration of remote magnetometry took place at the Observatorio del Roque de los Muchachos (ORM) in La Palma (Canary Island, Spain)\,\cite{Pedreros2018}. A continuous-wave laser based on a fiber Raman amplifier and second harmonic generation was used to deliver up to 20\,W of power onto the sky through a 30\,cm diameter projector\,\cite{Bonaccini2012}. The continuous beam was intensity modulated with an acousto-optic modulator and dithered at 150\,Hz, following the same pumping scheme and strategy as Kane et al \cite{Kane2018}. The receiver consisted of an off-the-shelf telescope of 35\,cm diameter and a photomultiplier tube delivering photon-count signal into three detection channels. Fifty-one frequency scans were successfully performed and the average Larmor frequency among these scans was found to be 260.4(1)\,kHz, corresponding to the field of 37200(10)\,nT, less than 0.5\,\% away from the WMM2015 prediction. By numerical fitting the magnetic resonances, it was found that two superimposed Lorentzians of different widths provided the best description of the resonances, see Fig.\,\ref{fig:Pedreros2018Resonance}. The noise analysis concluded that on the steepest slope of the narrower resonance, a sensitivity of 28\,nT/$\sqrt{\text{Hz}}$ could be achieved in the best case, in contrast to a sensitivity of 124\,nT/$\sqrt{\text{Hz}}$ on the steepest slope of the broader resonance, similar to the sensitivity found in the previous experiment. Numerical simulations suggested that the origin of the double-resonance feature can be attributed to atomic collisions in the mesosphere, the broader resonance arising due to atoms leaving and reentering the Doppler distribution (velocity-changing collisions) and the width of the narrower resonance due to spin-exchange collisions between Na and O$_2$.

Another demonstration carried out by the same group using the same infrastructure explored polarization modulation of the cw beam as a modulation technique \cite{Pedreros2018polarization}. In this scheme, the switching between left-handed and right-handed circular polarization at a rate close to the Larmor frequency allows to pump synchronously with the precession of atomic polarization during one  cycle. In this scheme, one benefits from a stronger signal as the intensity of the beam is not modulated. To modulate the beam polarization, an electro-optic modulator was used. The fact that it had no resonant cavity permitted to scan a broad range of frequencies, only limited by the ability of the driving electronics to supply the required high frequency high-voltage signal to reach full modulation. The scope of the experiment was limited to detection of magnetic resonance and determination of the Larmor frequency from the modulation-frequency scans. A median Larmor frequency of 260.0(3)\,kHz was obtained corresponding to the field of 37152(42)\,nT, consistent with what was found during the previous campaign in La Palma.

A fourth demonstration was carried out at the Lijiang Observatory (China), using a 18\,W continuous wave Raman-fiber-amplified frequency-doubled laser, employing the intensity-modulation approach of Ref.\,\cite{Fan2019remote}. The 1.8\,m receiver telescope was equipped with a narrow-band sodium filter, a CCD camera and a photon counter. The novelty in this experiment was the introduction of gated photon-counting detection implemented with a field-programmable gate array (FPGA) as a way to detect the signal from selected portions in the sodium layer and therefore as a mechanism to achieve altitude-resolved magnetic field measurements. The FPGA was used to control the intensity modulator and to acquire the signal during the counting-gate. The duration of the frequency sweep was only 250\,ms. The short scan helped removing low-frequency noise contributions such as wind turbulence, fluctuations of the laser power, and beam pointing. The signal during this short scan was too small to distinguish the resonance, so an accumulation of multiple scans was required. A Larmor frequency of 322.95\,kHz, corresponding to a mesospheric magnetic field of 46148\,nT was found. The  sensitivity was found to be 849\,nT/$\sqrt{\text{Hz}}$, limited by the small returned intensity and scintillation as the dominant source of noise. In a future version of the experiment, authors plan to introduce the lock-in detection technique, as used elsewhere.

\begin{figure}[htb]
\centering
\includegraphics[width=0.9\textwidth]{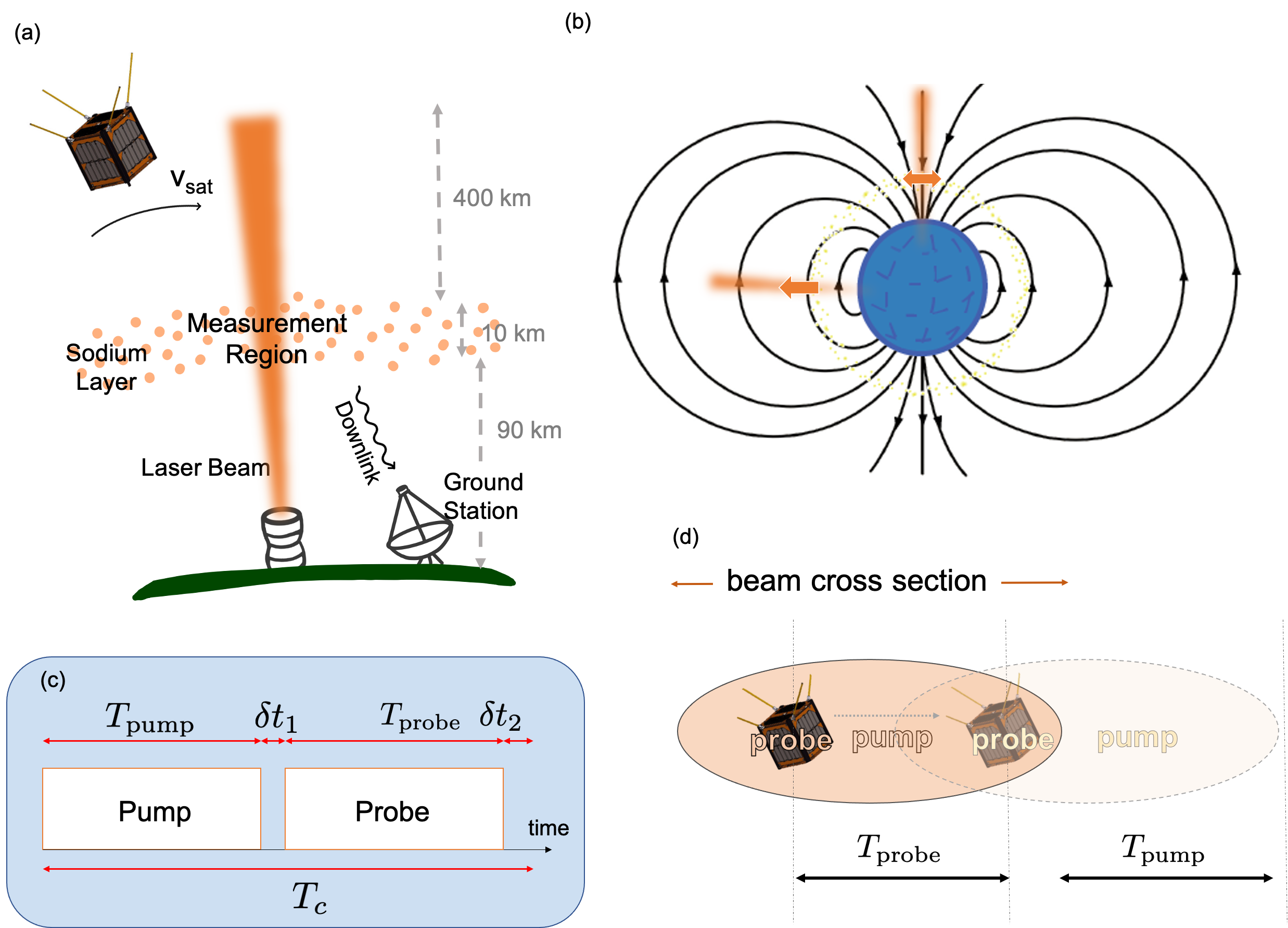}
\caption{(a) A schematic illustration of the experimental setup. (b) Measurements of Earth's magnetic field near the poles and at the equator. (c) Laser beam control sequence. (d) Laser beam cross section at the altitude of the satellite, showing pump and probe regions within the laser beam while tracking.}
\label{fig:remote-sat}
\end{figure}

The Mesospheric Optical Magnetometry project in Norway aims at investigating the so-called field-aligned (Birkeland) electric currents in the atmosphere using LGS magnetometry as a tool to sense local disturbances of the magnetic field in the auroral region. (Scientific motivations of this experiment are discussed in Sec.\,\ref{sec:LGS_applications}.) It is planned to adapt the existing lidar facility at the Arctic Lidar Observatory for Middle Atmosphere Research (ALOMAR) observatory in northern Norway, which is regularly used to measure temperature and wind profiles in the sodium layer. One of the challenges of this experiment is the reduced sensitivity arising from the small angle between the laser beam and the magnetic field when pointing at zenith. As discussed in Sec.\,\ref{sec:sat-mag}, see Fig.\,\ref{fig:remote-sat}(b), a possible solution can be using alignment- rather than orientation-based magnetometry. Alternatively, the capability of the lidar facility to steer the beam up to 30\degree\ zenith angle would allow pointing the beam in a more favorable direction to obtain better sensitivity \cite{Serrano2017}. The large area of the 1.8\,m diameter receiver telescope will be essential for maximizing the photon collection efficiency. 

\subsection{Summary and future directions}

Lock-in detection was one of the key elements for the success of the experiments. Otherwise, direct measurements should be fast enough to minimize intensity fluctuations due to scintillation and sodium-density fluctuations, at the cost of a low signal strength. As the time of flight of the pulses changes with the distance to the mesosphere for different elevation angles, the phase of the lock-in detection must be appropriately adjusted. 

While the agreement between the observed and simulated resonances (see, for example,  Fig.\,\ref{fig:Pedreros2018Resonance}) is generally satisfactory, it is not perfect. It could be improved with additional sensors to monitor the atmospheric turbulence, which determines the statistical intensity distribution of the laser beam in the mesosphere, and a sodium lidar, which allows to track changes in the vertical sodium distribution, would provide realistic input values to the physics models. As it is, since the relevant quantities like the Na density can vary during a night by a factor of 2-10, some discrepancies between observations and models are expected.

A high throughput receiver is essential to maximize the strength of the return signal. A telescope with a large collecting area is desirable, provided that the reflectivities of all surfaces are optimal. In practice, a smaller telescope could provide comparable efficiency, due to the minimal number of optical surfaces and the ease of maintaining high reflectivity of all components. Regarding laser technology, a boost in return flux could be obtained with higher-power lasers and the introduction of chirping technology \cite{PedrerosBustos:2020,hellemeier2022laser}. The basic idea here is that the use of high laser power desired for high photon return leads to the depletion of the resonant velocity group of atoms. Since the absorbed laser photons always impart a recoil in the direction of the laser beams, the atoms accumulate in a velocity group adjacent to the resonant one with the velocity directed away from the laser. If the laser frequency is chirped, this can lead to a noticeable improvement of the return and, correspondingly, the statistical sensitivity of remote magnetometry. 

% Table generated by Excel2LaTeX from sheet 'Sheet2'
\begin{table}[htbp]
 \def\arraystretch{1.3}
\begin{threeparttable}[b]
\begin{scriptsize}
  \centering
  \caption{Summary of mesospheric magnetometry experiments.}
    \begin{tabular}{l|c|c|c|c|c}
%    \toprule
\hline
        & \multicolumn{5}{c}{Site} \\[0.1cm]
\cline{2-6} 
    Parameters & Steward Observ. & ORM\tnote{1}   & ORM\tnote{1}   & Lijiang Observ. & ALOMAR\tnote{2}  \\
          & Arizona, USA & La Palma, Spain & La Palma, Spain & China & Observ., Norway \\[0.1cm]
    \hline
   Peak laser power & 3.8\,W   & 15\,W    & 12.5\,W  & 18\,W    & 10\,W \\[0.1cm]

 %   \midrule
 \hline
    Modulat. format & Pulsed, intensity & Pulsed, intensity & cw, polarization & Pulsed, intensity & Pulsed, intensity \\[0.1cm]
 %   \midrule
 \hline 
    Polarization & $\sigma^-$ & $\sigma^-$ & $\sigma^+ \leftrightarrow \sigma^-$ & $\sigma^-$ &  - \\[0.1cm]
 %   \midrule
 \hline
    Receiver diameter  & 1.55\,m  & 0.35\,m  & 0.35\,m  & 1.8\,m   & 1.8\,m \\[0.1cm]
 %   \midrule
 \hline 
    Larmor frequency  & 318.0\,kHz & 260.4(1)\,kHz & 260.0(3)\,kHz & 322.6\,kHz &  350\,kHz\tnote{3} \\[0.1cm]
 %   \midrule
 \hline
    Sensitivity & 162\,nT/$\sqrt{\text{Hz}}$   & 28\,nT/$\sqrt{\text{Hz}}$    & not stated & 849\,nT/$\sqrt{\text{Hz}}$   &  - \\[0.1cm]
 %   \midrule
 \hline
    $B$-field angle & 60\,\degree    & 66\,\degree\,to 90\,\degree & 80\,\degree    & 58\,\degree    & 138--168\,\degree \\[0.1cm]
 %   \midrule
 \hline
   Elevation  & 60\,\degree    & 51\,\degree\,to 75\,\degree & 64\,\degree    & 90\,\degree    & 60\,\degree \\[0.1cm]
 %   \midrule
 \hline
    Comments &   Ref.\,\cite{Kane2018}    & Ref.\,\cite{Pedreros2018}       &  Ref.\,\cite{Pedreros2018polarization}     &  Ref.\,\cite{Fan2019remote}     & Ref.\,\cite{Serrano2017}\\
    & & & & &  Planned expt. \\[0.1cm]
    \hline
    \hline
    \end{tabular}%
    \begin{tablenotes}
    \item [1] Observatorio del Roque de los Muchachos.
    \item [2] Arctic Lidar Observatory for Middle Atmosphere Research.
    \item [3] Expected Larmor frequency based on ground magnetic field data.
    \end{tablenotes}
  \label{table:MagExperiments}%
\end{scriptsize}
\end{threeparttable}
\end{table}%

\subsection{Application: geomagnetism} \label{sec:LGS_applications}

%\subsubsection{Geomagnetism}

Geomagnetic field is extensively monitored with a global network of magnetic observatories on the Earth surface \cite{Love:2013} and with magnetometers on board satellites \cite{Maus:2007,friis2006swarm}. In the past, vertical profiles were measured using rockets \cite{Singer:1950,Cahill:1956,Heppner:1958,Hutchinson:1961,Burrows:1965,Burrows:1971,Cloutier:1972}. Less frequently, aerial surveys are performed using magnetometers suspended from a helicopter or a plane flying a few kilometers above the surface, and typically cover a specific area of interest (group of islands, region, etc.) \cite{Garcia:2007,BlancoMontenegro:2007}. A review on the magnetometer technologies in aerospace applications can be found in Ref.\,\cite{Bennett:2021}. Scarce information exists between near-Earth ($\approx 2$~km) and space magnetic observations ($\approx 300$~km), and this lack of information translates into altitude inaccuracy of geomagnetic models and reduced knowledge of geologic and tectonic features at these spatial scales. 

The geomagnetic field is a continuum which connects different parts of Earth to each other and to nearby space. Also, the geomagnetic field has a broad spectrum of spatial scales, each of which is linked to different sources and processes \cite{Lowes:1974,Maus:2008b}. The geomagnetic field is time-varying from slow geological time scales to fast events connected to space weather and its influence in the ionosphere and magnetosphere. Therefore, continuously monitoring the geomagnetic field in time and space is crucial for understanding the complex interplay among the processes that generate and modify the geomagnetic field, as well as for practical applications like mineral and archaeological prospecting \cite{Love:2008}.

It is particularly important to understand the geophysical processes occurring in the lithosphere, which is the outermost solid layer of Earth comprising the crust and the upper mantle. Processes such as volcanic activity, motion of tectonic plates (leading to earthquakes), soil formation, and oceanic crust subduction, can be studied through the measurement of magnetic signatures enabling remote access to internal properties of Earth. These signatures are frequently associated with magnetic anomalies or small deviations from the main field. However, measurements from space lack high spatial resolution due to the long distance from lithospheric sources. Surface measurements are affected by local magnetic sources and insufficient for providing adequate spatial information in a continuous manner. Here, the altitude of magnetic field observations is a critical variable for high-resolution mapping, and for separating external magnetic fields from the lithospheric contribution \cite{Thebault:2010,Maus:2008}. For these reasons, monitoring the geomagnetic field at 85--100\ km altitude continuously and with spatial resolution could contribute to the mapping and understanding of the inner Earth. 

It is also important to have access to measuring the magnetic field at high altitude because it allows to monitor space weather as the main driver of the external field. Due to the proximity of the sodium layer to the D and E regions of the ionosphere (between 70 km to 120 km altitude) measuring the geomagnetic field in the mesosphere opens the possibility to map local current structures in the dynamo region \cite{Yamazaki:2017}. 

%Finally, monitoring the Earth magnetic field can contribute to mitigating risks associated with extreme events related to space-weather activity, like solar storms and their effect on electrical power grids. Over several decades, there have been unceasing efforts to establish electromagnetic precursors to earthquakes \cite{DeSantis:2017,Ghamry2021,Johnston:1997,Freund2011,Balasis2007} and volcano eruptions~\cite{Johnston2007,Uyeda2002} because prediction of such events is of importance for mitigating their potentially catastrophic consequences.

%In September 2016, scientists published a study that revealed a direct link between GPS blackouts of low-Earth-orbiting satellites and "thunderstorms" in the ionosphere. During the first two years of Swarm's operation their GPS connection was broken 166 times.[9] The high-resolution observations from the satellite helped to link these outages to ionospheric thunderstorms 300–600 km in the Earth's atmosphere.[10]
%\url{https://agupubs.onlinelibrary.wiley.com/doi/full/10.1002/2016SW001439}

%In December 2016, scientists announced that, by using data from the Swarm satellites, they had discovered a new feature in the Earth's outer core, a jet-stream of rapidly moving liquid iron moving at around 50 km per year.[11][12]
%\url{https://www.nature.com/articles/ngeo2859}

In fact, there is an increasing interest to employ LGS magnetometry as a tool to investigate the magnetic field disturbances in the mesosphere around the polar region \cite{Serrano2017}. The driver of this research is to advance the understanding of the ionospheric current system whose changes are connected to solar activity, gravity waves, and the interaction between the ionosphere and thermosphere system \cite{Eriksen2019}. The technique of LGS magnetometry has the potential to probe magnetic field disturbances produced by small-scale auroral electric currents in the E-region that can only be detected by a sensor nearby, where only sporadic research rocket measurements have access to.

\subsection{Directions for future progress}

After the successful demonstrations of LGS based magnetometry, there are many directions for refinement and improvement of the technique, including:
\begin{itemize}
    \item Development of compact systems for deployment on regional and global networks of magnetometers. Laser technology has evolved such that robust systems currently employed in astronomical observatories, could be also transformed into dedicated magnetometry systems for field deployment.   
    \item Increased sensitivity with readily available higher power guide star lasers (the power of the cw lasers currently used in LGS is approaching 100\,W).
    \item Development of methods to measure the $B$-field in polar regions (for example, by using alignment based magnetometry).
    \item Development of methods to sense gradients in the $B$-field. This can be done by ``synthetic gradiometry'', i.e., by combining readings from several LGS magnetometers or by toggling the LGS location on the sky. 
    \item Development of methods to get altitude resolution, likely with pulsed laser and gated receivers.
    \item Development of methods to sense during daytime, which would bring additional information about daylight phenomena and allow uninterrupted around-the-clock observation.
    \item Improving the sensitivity by increasing the return-photon signal via mirrorless lasing Sec.\,\ref{sec:mirrorless-lasing} or satellite-assisted techniques Sec.\,\ref{sec:sat-mag}.
\end{itemize}

\newpage

% ------------------------ End section LGS magnetometry -------------------------

 \clearpage
 % https://www.overleaf.com/project/60168abb91029f70755b6a9c
\section{Mirrorless-lasing magnetometry} \label{sec:mirrorless-lasing}

One of the possible detection modalities considered as part of the original REMAS project (Sec.\,\ref{sec:REMAS}) involved detection of amplified spontaneous emission (ASE) from excited Xe or Kr atoms. The potential advantage of such detection as opposed to detecting nearly isotropic fluorescence is that the ASE beam could be directional, with one of the preferential emission directions being back towards the source. Obtaining directional emission from atmospheric Xe and Kr is challenging, and, as far as we know was not accomplished in relation to REMAS. The basic idea was revived many years later in the context of LGS magnetometry, although on-sky demonstration has not yet been reported.

Although LGS-based magnetometry, see Sec.\,\ref{sec:LGS-mag}, is the state of the art of remote Earth-field magnetometry, the sensitivity of current scheme, based on nearly isotropic fluorescence, is limited by the scarce amount of detected light from the mesosphere, because the small solid angle (typ. $10^{-10}$) of detection. This strongly limits the magnetic field sensitivity to about $28\,\rm{nT}/\sqrt{\rm{Hz}}$, necessitating  measurement times over several minutes in order to achieve the desired levels of sensitivity with this technique \cite{bustos2018remote}. Thus, a natural improvement of LGS-based techniques for remote magnetometry relies on increasing the return flux of the LGS. In addition to improving the magnetic sensitivity, increasing the return flux would have a high impact in astronomy because the performance of adaptive optics systems increases with the return flux, and the requirements are stringent in view of the upcoming generation of extremely large telescopes. Increasing the return flux could also enable daylight operation of LGS based instrumentation.

One way to address this issue is to ``force" the atoms to radiate back toward the source. The use of mirrorless lasing \cite{Allen1970} for increasing the LGS return flux was suggested in \cite{Pedreros2018PLGS}. It was indeed demonstrated, in laboratory conditions, that highly directional emission of infrared light in the backward direction can be generated from sodium vapors excited by two cw lasers \cite{akulshin2018continuous}, see Sec.\,\ref{sec:mirrorless-lasing-concept}.  Later, these studies were extended to remote detection of the sodium ground-state free-precession under the influence of an external magnetic field by the generated infrared light \cite{Zhang2021_Standoff}, see Sec.\,\ref{sec:mirrorless-lasing-realization}. The latter is promising for remote sensing of Earth magnetic field as the technique enables scalar magnetometry in the Earth-field range without the need of calibration, see Sec.\,\ref{sec:mirrorless-lasing-applications}. A work-in-progress modelling of mirrorless lasing is presented in Sec.\,\ref{sec:mirrorless-lasing-theory}. Scaling of this technique to the sky is being investigated, see Sec.\,\ref{sec:mirrorless-lasing-scaling}.

\subsection{The concept}\label{sec:mirrorless-lasing-concept}
 
The results reported by in Ref.\,\cite{akulshin2018continuous} originated from the idea of inducing laser-like emission from the sodium layer, which due to its high directionality in the backwards direction, could increase the return flux by several orders of magnitude with respect to traditional sodium laser guide stars. Although the general physical mechanisms of ASE generation were understood, other effects were found during the experiments, for example the spiking behavior of ML \cite{akulshin2021intensity} and the dependence of the strength of coherent emission on the local magnetic field. The latter was recently used to perform stand-off magnetometry from a sodium vapor cell \cite{Zhang2021_Standoff} as discussed in the following section.

\subsection{Experimental realization}
\label{sec:mirrorless-lasing-realization}
%\EK{Rui is writing this.}

\begin{figure}
\centering
\includegraphics[width=1.0\textwidth]{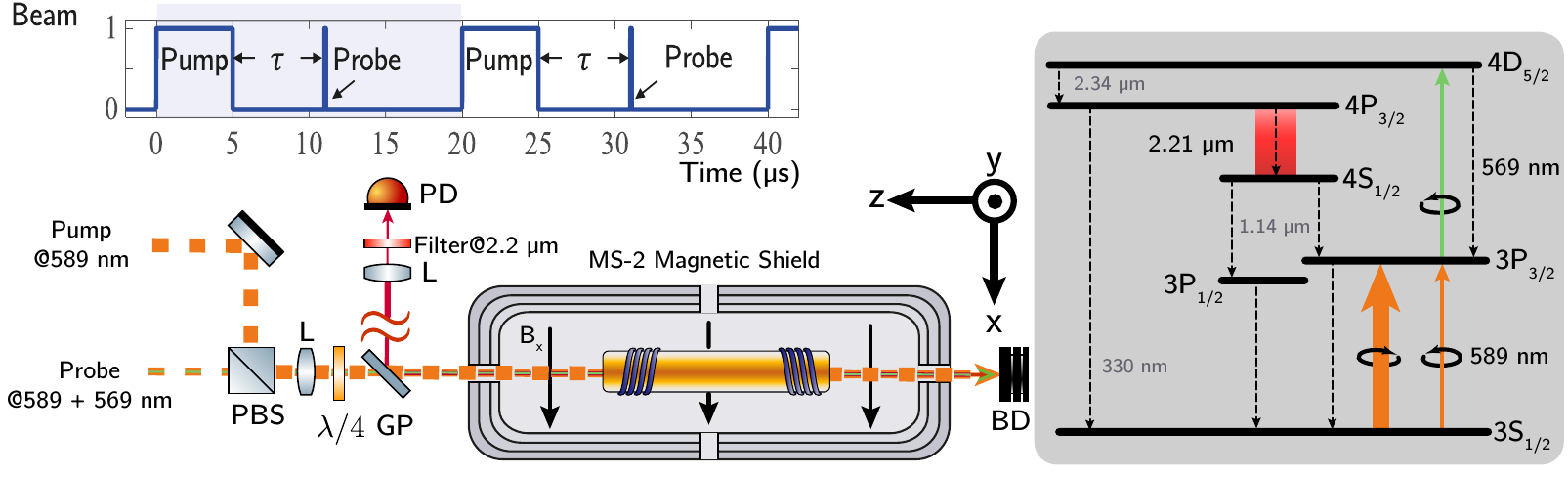}
\caption{Experimental setup for ML-based stand-off magnetometry. Blue shading on the upper left represents a measurement period. PD -- photodiode; PBS -- polarizing beam splitter; GP -- glass plate; L -- lens; BD -- beam dump. The right inset shows the energy levels of sodium }
\label{fig:remote-mag-ML}
\end{figure}

The energy level scheme used in Ref.\,\cite{akulshin2018continuous} to demonstrate Continuous mirrorless lasing in sodium vapor is shown in Fig.\,\ref{fig:remote-mag-ML}. Lasing was based on amplified $2.21\,\mu{\rm{m}}$ spontaneous emission between the $4\text{P}_{3/2}$ and $4\text{S}_{1/2}$ states. To achieve this, the atoms in the $3\text{S}_{1/2}$ ground level were pumped successively to the $4\text{D}_{5/2}$ state with a $589\,{\rm{nm}}$ light beam resonant with the $3\text{S}_{1/2}$ to $3\text{P}_{3/2}$ transition and a $569\,{\rm{nm}}$ light beam resonant with the $3\text{P}_{3/2}$ to $4\text{D}_{5/2}$ transition. These two co-propagating light beams were spatially overlapped and were set to the same circular polarization to enhance the overall pumping rate. Atoms in the $4\text{D}_{5/2}$ state decay into the $4\text{P}_{3/2}$ state, resulting in population inversion between the $4\text{P}_{3/2}$ and $4\text{S}_{1/2}$ states in an elongated interaction region defined by the light beams. Such population inversion provides gain for the $2.21\,\mu{\rm{m}}$ spontaneous emission and leads to $2.21\,\mu{\rm{m}}$ mirrorless lasing both in the forward and backward directions. The backward-directed radiation can be used for remote magnetometry, as the amount of population inversion and the intensity of subsequent mirrorless lasing are dependent on the ground-state spin polarization, and thus are influenced by the magnetic field.

\begin{figure}[tb]
\centering
\includegraphics[width=1.0\textwidth]{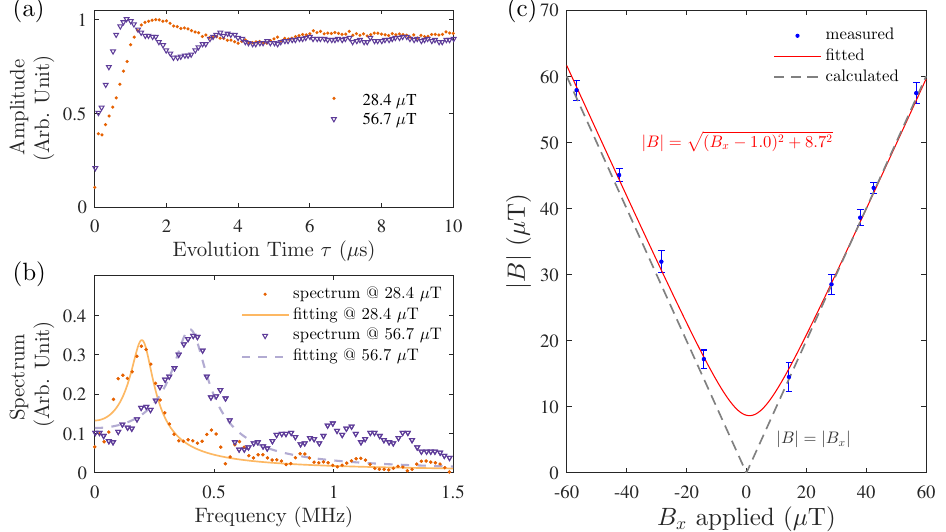}
\caption{ Experimental results of ML-based stand-off magnetometry~\cite{Zhang2021_Standoff}.
    \textbf{(a)} Free-precession signals. \textbf{(b)} The amplitude spectra of the free precession signals shown in (a) and its Lorentzian fittings. \textbf{(c)} Measured magnetic fields at different bias magnetic field applied.
 }
\label{fig:remote-mag-ML-Results}
\end{figure}

A challenge for extending mirrorless lasing detection to high-sensitivity magnetometry is that the strong light required for generation of mirrorless lasing inevitably disrupts the ground-state spin precession, which is incompatible with continuous readout of spin polarization commonly used in atomic magnetometers.
To overcome this challenge, a three-stage method based on the principle of ``free'' Larmor precession of the ground state atomic spins was adopted \cite{Zhang2021_Standoff}. The three stages are: 1. Optical pumping of the ground-state spin polarization with a 589\,nm pump beam tuned to the sodium D$_2$-line ($3\text{S}_{1/2} \rightarrow 3\text{P}_{3/2}$ transition); 2. Free evolution of the atomic spins ``in the dark''; 3. Readout of the atomic polarization via mirrorless lasing, the intensity of which depends on the ground-state spin polarization.
The measurement is repeated with different evolution times, and the combined result reconstructs the full evolution of spin polarization, which is a damped oscillation with a central frequency at the Larmor frequency, see Fig.\,\ref{fig:remote-mag-ML-Results}(a).
Measuring the central frequency enables scalar magnetometry that does not require calibration or precise knowledge of multiple experimental parameters in mirrorless lasing, see Fig.\,\ref{fig:remote-mag-ML-Results}(b) and Fig.\,\ref{fig:remote-mag-ML-Results}(c).

\subsection{Applications} \label{sec:mirrorless-lasing-applications}

%\EK{This subsection is not finished.}

% The dependence of ML intensity on magnetic field was utilized by Zhang et al\,\cite{Zhang2021_Standoff}, where a mirrorless-lasing-based remote magnetometer was demonstrated in a vapor cell based table-top experiment. The sodium ground-state free precession under the influence of an external magnetic field was observed by recording the intensity of the backward-directed ML, and used to demonstrate absolute and calibration-free stand-off magnetometry, see Sec.\,\ref{sec:mirrorless-lasing-realization}. 
As most of the experimental apparatus can be remote from the atomic vapor sensing the magnetic field, this mirrorless lasing-based stand-off magnetometry provides opportunity for hazardous environment monitoring or remote geophysical research. For example, it can be operated in harsh environments near nuclear reactors or particle accelerators, where it is impossible to place conventional sensors or electronics. 

In the case of vapor cell mirrorless lasing magnetometry, the sensitivity could, in principle, be further improved if the vapor cells are filled with buffer gas such as He, N$_2$ or Ne to extend coherence times as the atoms would spend more time in the light-beam area. This is a subject of ongoing research.

Apart from using vapor cells or the mesospheric sodium, sodium vapor can also be produced in plasma near the surface of salt water (or similar environments). Magnetometry with sodium atoms in such plasma was recently demonstrated \cite{ding2022magnetometry} and this approach can potentially be combined with mirrorless lasing, another topic of current research.

\subsection{Theory}\label{sec:mirrorless-lasing-theory}

The problem of directed mirrorless emission has been studied by various authors in the decades since the invention of the laser \cite{Allen1973,Garrison:88,nastishin2013optically}. In the context of remote sensing, it is typical not to make a distinction between amplified spontaneous emission and mirrorless lasing, and we follow this approach here as well, even though one may discuss possible differences based on, for instance, the second-order correlation function of the light.

Below we briefly outline our work-in-progress estimates relevant for atmospheric remote magnetometry. A similar discussion can be found in Ref.\,\cite{hickson2021can}.

\begin{figure}
    \centering
    \includegraphics[width=0.8\textwidth]{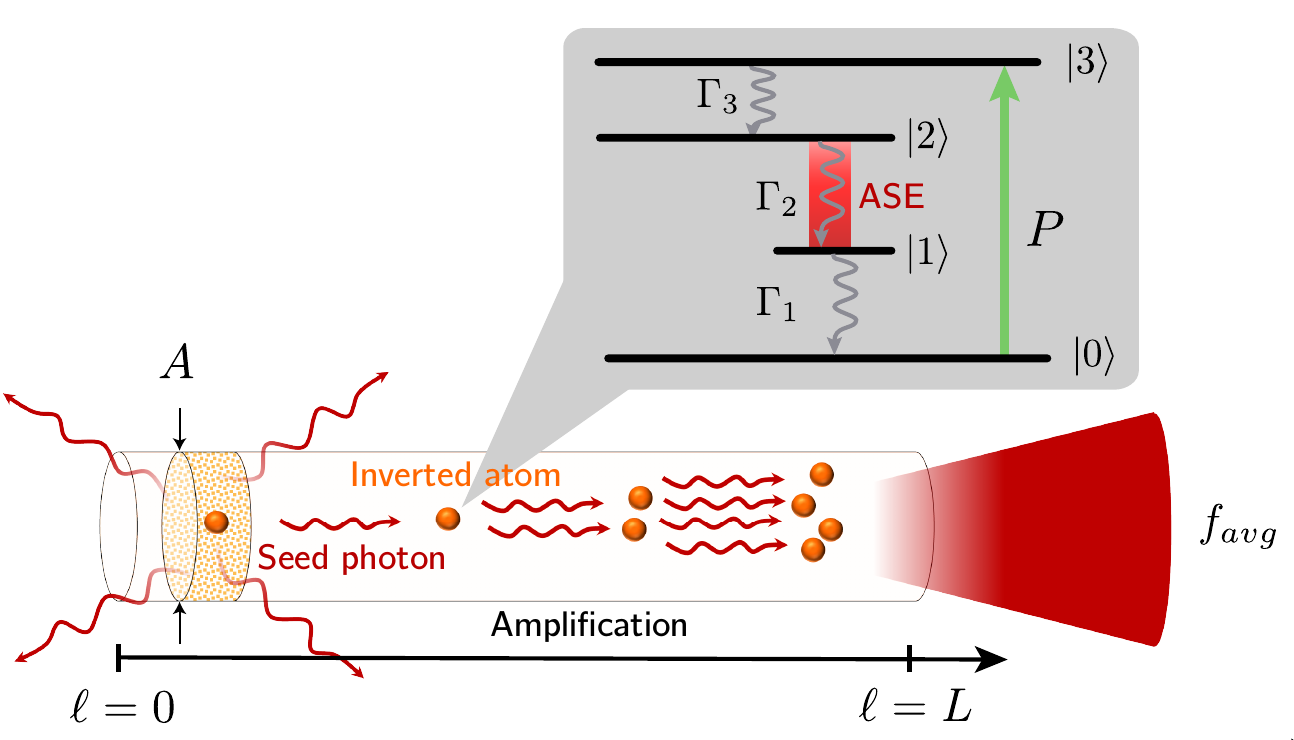}
    \caption{Mirrorless lasing in a pencil-shaped sample. The inset shows a simplified level and transition scheme (see text).}
    \label{fig:MLTheory}
\end{figure}

Consider an atomic medium with a ground state $\ket{0}$ and three other states that decay in a cascade $\ket{3} \to \ket{2} \to \ket{1} \to \ket{0}$ (see the inset in Fig.\,\ref{fig:MLTheory}). Atoms are pumped from the ground state to the state $\ket{3}$ at a rate $P$. If the effect of reabsorbed light on the level populations is negligible (an assumption examined below), the steady-state population densities $n_{0,1,2,3}$ of the states can be deduced from
\begin{equation}\label{eq:four-level-steady-state-eqs}
\begin{gathered}
    n_0 + n_1 + n_2 + n_3 = n_t\,,\\
    \Gamma_2 n_2 - \Gamma_1 n_1 = 0\,,\\
    \Gamma_3 n_3 - \Gamma_2 n_2 = 0\,,\\
    P n_0 - \Gamma_3 n_3 = 0\,,
\end{gathered}
\end{equation}
%\EK{In the case of coherent pumping shouldn't it be $P(n_0-n_3)$?} 
where $n_t$ is the total population density and $\Gamma_i$ are the spontaneous decay rates with all other decay branches being negligible. Assuming high enough pumping rate (another assumption examined below), $P$ drops out and we find 
\begin{equation}
\begin{gathered}
    n_1 = \frac{n_t}{1+\Gamma_1/\Gamma_2+\Gamma_1/\Gamma_3}\,,\\
    n_2 = \frac{n_t}{1+\Gamma_2/\Gamma_1+\Gamma_2/\Gamma_3}\,.
\end{gathered}
\end{equation}
This gives for the inversion density $n_i = n_2 - n_1$ on the $\ket{1}\rightarrow\ket{2}$ transition 
\begin{equation}
    n_i = \frac{\Gamma_1/\Gamma_2 - 1}{1 + \Gamma_1/\Gamma_2 + \Gamma_1/\Gamma_3}n_t\,.
\end{equation}
We see that population inversion (positive $n_i$) can be obtained if $\Gamma_1 > \Gamma_2$, i.e., if atoms decay out of state $\ket{1}$ faster than they are replenished by decay from state $\ket{2}$.

Now consider a column of such atoms with cross-section $A$ and length $L$. Photons are spontaneously emitted on the $\ket{2}\to\ket{1}$ transition at a rate $\Gamma_2 n_2$. For simplicity, we assume that the photons are emitted isotropically. A photon resonant with the $\ket{1}\to\ket{2}$ transition that travels a distance $d\ell$ through the medium has a probability $n_1\sigma d\ell$ to be absorbed and $n_2\sigma d\ell$ to stimulate emission, for a net emission probability $n_i\sigma d\ell$, where $\sigma$ is the resonant absorption cross section for the transition. ($\sigma$ is proportional to the square of the transition wavelength, with numerical factors depending on the angular momenta and branching ratio involved in the transition.)  If the transition is Doppler broadened with width $\Gamma_D$, only a fraction $\Gamma_2/\Gamma_D$ of the photons will be resonant, so the average emission probability over all photons is $d\ell/\ell_i$, where $\ell_i=\Gamma_D/(n_i\sigma\Gamma_2)$ is the ``net emission length'' -- the travel distance through the atomic medium required for a photon to induce at least one net emission on average. Considering any straight-line path through the medium, we write $L_\Omega(\ell)$ (in units of photons/s/sr/m$^2$) for the radiance of photons traveling along the path, i.e., the number of photons per unit time passing through an area element $dA$ containing the point at distance $\ell$ along the path within a differential solid angle $d\Omega$ containing the path direction. Including the spontaneous and stimulated emission terms from above, the change in radiance along such a straight-line path is described by 
\begin{equation}
    \frac{dL_\Omega(\ell)}{d\ell} 
    = \frac{L_\Omega(\ell)}{\ell_i} + \frac{\Gamma_2n_2}{4\pi}\,.
\end{equation}
Integrating over the path, we have
\begin{equation}\label{eq:ASE_radiance}
    L_\Omega(\ell) = \frac{\ell_i n_2\Gamma_2}{4\pi}(e^{\ell/\ell_i} - 1)\,.
\end{equation}
If we take the detector to be far from, and on-axis with, the column, we need consider only photons that travel nearly lengthwise down the column with path length $L$ in the medium. Thus, integrating over the area $A$, we find for the photon radiant intensity $I_\Omega$ (in photons per steradian) at the detector
\begin{equation}\label{eq:ASE_intensity}
    I_{\Omega, \mathrm{det}} 
    = \frac{A\ell_i n_2\Gamma_2}{4\pi}(e^{L/\ell_i} - 1)
    = \frac{A}{4\pi\sigma}\frac{\Gamma_1\Gamma_D}{\Gamma_1-\Gamma_2}(e^{L/\ell_i} - 1)\,.
\end{equation}
If the detector is off-axis from the column, a geometrical factor will arise due to the fact that photons arriving at the detector have traveled different lengths through the active medium.

For an optically thin medium, $L/\ell_i\ll1$, the radiant intensity (\ref{eq:ASE_intensity}) reduces to 
\begin{equation}
    I_{\Omega, \mathrm{det}} \approx \frac{A L n_2\Gamma_2}{4\pi}\,,
\end{equation}
i.e., the photons due to spontaneous emission only, and the intensity increases linearly with  $L/\ell_i$. On the other hand, above the optical thickness threshold $L/\ell_i>1$, Eq.~(\ref{eq:ASE_intensity}) indicates that the intensity approaches exponential growth. This threshold condition gives the minimum atomic density needed for mirrorless lasing. If we drop the assumption of very high pumping rate $P$, the optical thickness also becomes a function of $P$, leading to interdependent thresholds for density and pumping rate.

The exponential increase of emission with optical thickness cannot continue forever. If the photon irradiance (photon flux per unit area) $E$ becomes large enough to perform significant pumping on the $\ket{1}\to\ket{2}$ transition, the $\ket{1}$ and $\ket{2}$ populations will start to equalize, causing saturation. Saturation occurs at an irradiance on the order of 
\begin{equation}\label{eq:saturation_irradiance}
    E \sim \frac{\Gamma_D}{\sigma}\,.
\end{equation}
The highest irradiance occurs at the end of the column. We can find the irradiance there by integrating the radiance (\ref{eq:ASE_radiance}) over solid angle. For a sufficiently long and narrow column ($A\ll L^2$) and large enough optical thickness, the intensity will be dominated by the photons that have traveled on the paths of length $L$ that begin at the other end of the column. These paths occupy a solid angle of $A/L^2$, so the integration is equivalent to multiplying (\ref{eq:ASE_radiance}) by $A/L^2$ and setting $\ell=L$: 
\begin{equation}
    E_\mathrm{end} 
    % =\int d\Omega\frac{\ell_i n_2\Gamma_2}{4\pi}(e^{\ell(\Omega)/\ell_i} - 1)
    =\frac{A}{L^2}\frac{\ell_i n_2\Gamma_2}{4\pi}(e^{L/\ell_i} - 1)\,.
    % =\frac{1}{L^2}\frac{d\tilde{\Phi}_d}{d\Omega}
\end{equation}
Comparing with Eq.~(\ref{eq:ASE_intensity}), we see that the irradiance at the end of the column is related to the radiant intensity at the detector by
\begin{equation}\label{eq:irradiance_vs_intensity}
    E_\mathrm{end} 
    % =\int d\Omega\frac{\ell_i n_2\Gamma_2}{4\pi}(e^{\ell(\Omega)/\ell_i} - 1)
    % =\frac{\ell_i n_2\Gamma_2}{4\pi}(e^{\ell/\ell_i} - 1)\frac{A}{L^2}\,.
    =\frac{I_{\Omega, \mathrm{det}}}{L^2}\,.
\end{equation}
Thus, equating (\ref{eq:saturation_irradiance}) and (\ref{eq:irradiance_vs_intensity}), the ASE radiant intensity at the detector will begin to saturate at 
\begin{equation}
    I_{\Omega, \mathrm{det}} \sim \frac{L^2\Gamma_D}{\sigma}\,,
\end{equation}
which, from Eq.~(\ref{eq:ASE_intensity}), corresponds to an optical thickness of on the order of
\begin{equation}
    L/\ell_i \sim \ln\left(4\pi\frac{L^2}{A}\frac{\Gamma_1-\Gamma_2}{\Gamma_1}\right)\,.
\end{equation}

\begin{figure}
    \centering
    \includegraphics[width=4.5in]{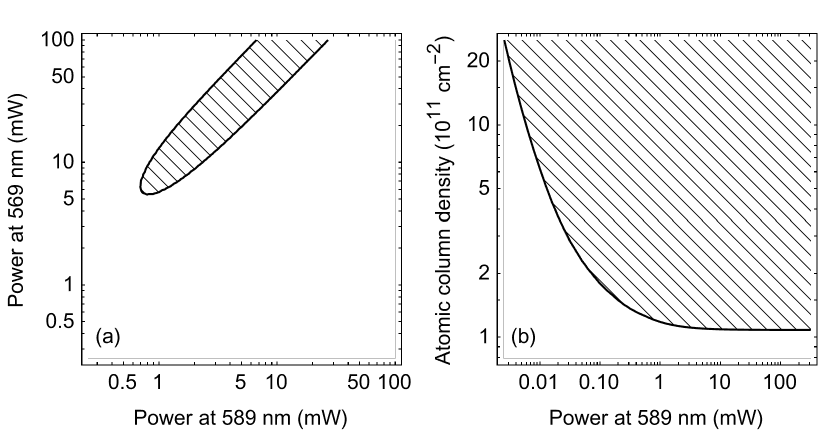}
    \caption{(a) Estimated above-threshold mirrorless lasing region (shaded) as a function of 589\,nm and 569\,nm pump power at a fixed atomic column density ($1.2\times10^{11}$\,cm$^{-2}$) for the conditions of the experiment in Ref.~\cite{akulshin2018continuous}. (b) Above-threshold region as a function of 589\,nm pump power and column density, taking 569\,nm pump power to be eight times 589\,nm power.}
    \label{fig:sodium-cell-ASE-threshold}
\end{figure}

We can apply an analysis of this sort to estimate the threshold pump laser power and atomic density for the experimental realization of mirrorless lasing in sodium discussed in Sec.\,\ref{sec:mirrorless-lasing-realization} and Ref.\,\cite{akulshin2018continuous}. The sodium system is more complicated than the four-level system considered above, and requires two pump laser fields (Fig.\,\ref{fig:remote-mag-ML}). We form the equations for the atomic steady state [analogous to Eq.~(\ref{eq:four-level-steady-state-eqs})], including the Zeeman substructure for each level, but neglecting hyperfine structure for simplicity. We also include the effect of the atomic Doppler shifts by dividing the atomic velocity distribution into velocity groups and solving for each group separately. Using the reported values for the laser beam diameters and the temperature of the atomic medium \cite{akulshin2018continuous}, we can solve numerically to find the lasing threshold (optical thickness equal to one) as a function of laser power and atomic column density. Figure~\ref{fig:sodium-cell-ASE-threshold}(a) shows the pump powers above threshold (shaded region) for a fixed value of column density. We see that the optimal power in the 569\,nm pump field is proportional to that of the 589\,nm field. Varying the power of the two pump fields proportionally, we plot the region above threshold as a function of pump power and column density in Fig.\,\ref{fig:sodium-cell-ASE-threshold}(b). The estimated threshold values shown here are around an order of magnitude smaller than those reported in Ref.~\cite{akulshin2018continuous}. For example, the experimental threshold column density was found to be about $3\times10^{12}$\,cm$^{-2}$ for 30\,mW pump power at 589\,nm and 12\,mW at 569\,nm, whereas our estimate from Fig.\,\ref{fig:sodium-cell-ASE-threshold}(b) is less than $2\times10^{11}$\,cm$^{-2}$ for high enough pumping rates. This may be partially explained by simplifications that we made in the model, including neglecting hyperfine structure.

% \DB{Some things to consider}
% \begin{itemize}
%     \item We need to explain that we are assuming that we can have laser power as high as we need; can we still have lasing?
%     \item We should probably say explicitly that, in this context, by "laser" we mean any directional beam. The g$^{(2)}$ considerations are not taken into account here
%     \item The Na pencil does not go to the Earth, only 10 km; we need to take that into account, or just mention this
%     \item Connect to Sec. 5.6 or restructure the logic a bit
%    % \item Manu wants to add a few references about the theory work along these lines
% \end{itemize}

% \subsection{What species can be used for mirrorless-lasing magnetometry?}

% A directed light beam coming down towards the ground from an atmospheric species used for remote magnetic sensing would be highly desirable to avoid the loss of most of the emitted photons in the case of spontaneous emission. As discussed above, this could be accomplished with amplified spontaneous emission or mirrorless lasing. The requirement for this is that one should be able to produce sufficient gain on the transition in question. This translates into the requirement of creating population inversion in an optically thick sample. The question of which species present in the atmosphere can be used for this was recently addressed in Refs.\,\cite{hickson2021can,Yang2021atomic}, see also Sec.\,\ref{sec:mirrorless-lasing}. 

\subsection{Scaling to the sky}\label{sec:mirrorless-lasing-scaling}

\begin{figure}
    \centering
    \includegraphics[width=2.5in]{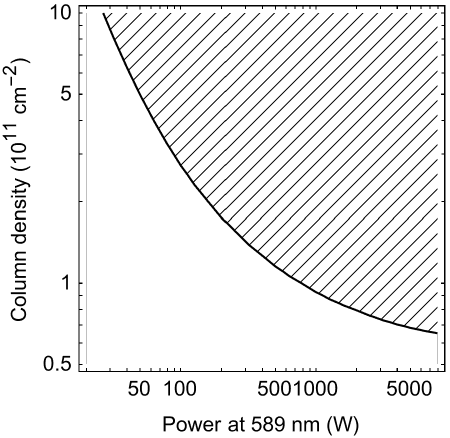}
    \caption{Estimated above-threshold mirrorless lasing region (shaded) for the mesospheric sodium layer as a function of 589 nm pump power and sodium column density, taking 569\,nm pump power to be eight times 589\,nm power.}
    \label{fig:sodium-layer-ASE-threshold}
\end{figure}

In analyzing a potential mirrorless-lasing experiment in the mesospheric sodium layer, there are a number of complicating factors relative to a vapor-cell experiment such as described in Ref.\,\cite{akulshin2018continuous}. These include collisions between the sodium atoms with nitrogen and oxygen molecules, which cause changes in atomic velocity, spin randomization, and quenching, the velocity-changing effect of radiation pressure due to the pump light, and spin precession due to the geomagnetic field \cite{Holzloehner2010AA}. However, if we neglect such complications, we can apply the same estimate discussed at the end of Sec.\,\ref{sec:mirrorless-lasing-theory} for the vapor-cell experiment to the mesospheric case. For the scheme with ASE at 2.2\,$\mu$m, this requires adjusting only the values of the laser-beam diameters, which we take to be 1\,m for the sodium-layer experiment, and the temperature of the medium, which is an average of 185\,K in the mesosphere. The resulting threshold curve is shown in Fig.\,\ref{fig:sodium-layer-ASE-threshold}. The much larger laser beam diameter means that higher pump power is required than for the vapor cell (note that the light power in Fig.\,\ref{fig:sodium-layer-ASE-threshold} is given in watts, as opposed to milliwatts for Fig.\,\ref{fig:sodium-cell-ASE-threshold}). The larger difficulty, though, is that the natural column density of the sodium layer is approximately $4\times10^9$\,cm$^{-2}$, more than an order of magnitude below the (likely optimistic) estimated threshold column density shown in Fig.\,\ref{fig:sodium-layer-ASE-threshold}, and there is no way to control this density, short of extreme measures such as seeding the mesosphere with sodium. 

Alternative pumping schemes involving ASE on other transitions can be more favorable. In particular, longer wavelength transitions have larger absorption cross sections (proportional to $\lambda^2$) and narrower Doppler widths (inversely proportional to $\lambda$), which tends to increase optical thickness. However, differences in lifetimes and branching ratios may cancel out these gains. For example, an estimate for a scheme involving 10.8\,$\mu$m ASE on the 6S$_{1/2}$ to 6P$_{3/2}$ transition of Na indicates that the column density requirement could be a factor of approximately three lower than for the estimate shown in Fig.\,\ref{fig:sodium-layer-ASE-threshold}.

A survey of all possible transitions in atoms and ions in the upper atmosphere that require either one- or two-step excitation was performed in Ref.~\cite{Yang2021atomic}. It was concluded that none of the metallic species present (Na, Fe, Mg$^+$, Si$^+$, Ca$^+$, and K) possesses a transition that could produce an optical thickness significantly greater than one, and so atmospheric mirrorless lasing in these species was ruled out. (The 10.8\,$\mu$m transition mentioned above provided the highest potential optical thickness.) On the other hand, atmospheric O, N, and N$^+$ were found to have transitions viable for ML, although excitation schemes would be challenging, as the excitation wavelengths are outside the atmospheric transmission window. 

A possible approach to reducing the column density requirement is to perform ``snowplowing'' of the atomic velocity distribution using frequency-chirped pump light, thus reducing the Doppler width of the mesospheric sodium. This type of technique has been studied \cite{PedrerosBustos:2020} and demonstrated \cite{hellemeier2022laser} as a means of combating the hole-burning effect of radiation pressure. However, a much more radical modification of the atomic velocity distribution than was shown in these studies would be needed in order to significantly reduce the threshold column density for mirrorless lasing.

\clearpage

\section{Satellite-assisted magnetometry of mesospheric sodium} \label{sec:sat-mag}
\subsection{The concept}
Satellite-assisted magnetometry was proposed as an alternative approach to address the relatively limited sensitivity and operating schemes of fluorescence based mesospheric Na magnetometry as discussed in Sec.\,\ref{sec:LGS-mag}. Motivated by recent advances in small, inexpensive satellite platforms such as CubeSats \cite{spacenews,centennial1}, this proposal suggests the direct detection of laser photons by employing a miniaturized trackable satellite equipped with a polarimeter. After optically pumping the sodium atoms in the mesosphere, the Larmor frequency could be measured by detecting the magneto-optical rotation or the transmission of the polarized probe-laser light with the photodetector onboard the satellite.

\subsection{Possible protocols and sensitivity}
The proposed system involves a ground-based telescope that launches pump and probe beams, and a trackable miniaturized satellite launched into low Earth orbit (LEO) at an altitude of 500\,km (this choice is somewhat arbitrary; we use this value for estimates). Once the satellite is within the laser spot, pumping is performed for about a millisecond (several atomic sodium ground-state spin relaxation times). The pump beam could either be modulated (in amplitude, frequency, or polarization) synchronously with the Larmor precession or be a strong continuous light pulse. The pumping interval is followed by optical probing of comparable duration when polarization rotation is monitored. Each pump-probe measurement cycle takes about 2\,ms and can be repeated as long as the satellite is visible within the laser beam. A wavelength of 589\,nm and a launch telescope diameter of 30\,cm result in a beam spot size of 0.5\,m at the Na layer (2.4\,m at the altitude of the satellite). With a typical guide star laser power of 20\,W, the sodium saturation parameter is on the order of $10^2$. This means that there is enough light to polarize the sodium atoms, and that the probing power needs to be attenuated to avoid repolarization. Alternatively, the beam area could be increased by two orders of magnitude so that efficient pumping can be achieved without introducing depolarization effects upon probing. 

In general, the launched laser beam makes an arbitrary angle with the Earth’s magnetic field. However, one can identify two extreme cases, namely measurements near the poles, where the magnetic field lines run parallel to the laser beam, and at the equator, where they are perpendicular to each other. The field-dependent magneto-optical rotation can be estimated given the Na absorption length and decay rate. In this measurement, we expect the sensitivity to be limited by photon shot noise, which is solely determined by the total number of photons $N_p$ that the satellite could detect. Assuming an on-board detector with a diameter of 10\,cm, $N_p$ is approximately \SI{1.0e15}{\second^{-1}}, leading to a sensitivity of $4.9\,\rm{pT}/\sqrt{\rm{Hz}}$ when measuring at the equator, and half that value at the poles.

% \begin{equation}
% \varphi_{1P} \approx \frac{\alpha}{2}\cdot\sin(2g_F\mu_B B t + \psi)\cdot e^{-\gamma_0 t},
% \label{eq:phi1P}
% \end{equation}
% and
% \begin{equation}
%     \varphi_{1E} \approx \frac{\alpha}{2}\cdot\sin(g_F\mu_B B t + \psi)\cdot e^{-\gamma_0 t}.
%     \label{eq:phi1E}
% \end{equation}
% where $\alpha$ is approximately 4\% given by the optical absorption  of the sodium layer, $g_F$ is the Landé $g$-factor for ground state Na, $\mu_B$ is the Bohr magneton, $\psi$ is the optical pumping phase, and $\gamma_0$ is the spin relaxation rate in the mesosphere of typically about $1/(250\,\mu \text{s})$ \cite{Pedreros2018}.

%%propagation time issues
%%magnetic field variations across the Na layer
% \subsection{Possible experimental realizations}

% \DB{Something is missing here?}\TD{I think everything from the origianl manuscript was summarized here? Thie subsection may have been added in the begining before the actual writing}

\clearpage
\section{Air lasing and its application to magnetometry} \label{sec:Air_Lasing}

Processes resembling mirrorless lasing in Na discussed in Sec.\,\ref{sec:mirrorless-lasing} can also occur with more common species occurring in the atmosphere. In fact,
a process of stimulated emission in which one of the naturally occurring species in the atmosphere turns into the active medium of a laser is referred to as ``air lasing''. This became possible with the advent of powerful pulsed lasers and remains an active area of current research \cite{polynkin2018air}.

A large body of work studies remote sensing diagnostics to monitor combustion processes in gas mixtures, see \cite{Alden2023}. In the context of remote atmospheric sensing as discussed above, directionality and coherence of amplified spontaneous emission boost the signal at the receiver. For combustion studies, lasing of some of the constituents is often more of an unwanted effect since interference effects have to be taken into account that often reduce spatial resolution. 

Air lasing was demonstrated with many different schemes and gain media. The paper describing the first experimental demonstration of amplified spontaneous emission in air recognized the potential for remote sensing \cite{Luo_2003_airlasing1,hemmerpnas2011}. In 2003 Luo et al. observed an exponential increase in the amount of back-scattered fluorescence from N$_2$ molecules that were subjected to an intense 42\,fs laser pulse. The high laser intensity of up to $5\times10^{13}$\,W/cm$^2$ creates a so-called filament in nitrogen gas, i.e. a low-density plasma column left behind a continuous series of self-foci as a result of the dynamic interplay of nonlinear effects. The amount of back-scattered light associated with transitions in molecular nitrogen as well as molecular nitrogen ions scaled exponentially with the length of the filament: a clear sign of amplification in the backward-scattered fluorescence.

\begin{figure}
    \centering
    \includegraphics[width=0.8\textwidth]{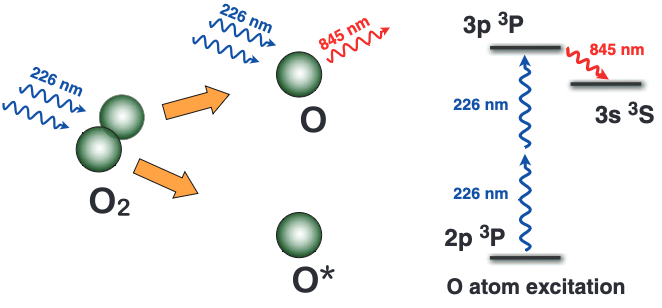}
    \caption{Schematic of the ``air-lasing'' experiment reproduced from Ref.\,\cite{dogariu2011high-gain}. Ultraviolet light at 226\,nm dissociates oxygen molecules and excites oxygen atoms from the ground state via two-photon absorption. Population inversion and mirrorless lasing occur at the infrared 845\,nm transition.}
    \label{fig:dogariu2010}
\end{figure}

In 2010, Arthur Dogariu et al. experimentally demonstrated high-gain backward lasing in air \cite{dogariu2011high-gain}. The experiment used a 226\,nm, 100\,ps pulsed laser to first dissociate molecular oxygen and then pump the resulting atomic oxygen in a two-photon process into an excited state, see Fig.\,\ref{fig:dogariu2010}. Optical gain and lasing are then observed at 845\,nm. 
%\DB{Can you add what transition this is? How about a diagram of the energy levels and transitions involved?} \EK{I made a quick thing (Fig.\ref{fig:dogariu2010}). Needs improvements and a caption.} 
Further measurements confirmed a gain of 62\,cm$^{-1}$ for this wavelength, the directionality via a measured divergence of 40\,mrad consistent with diffraction limited lasing from the cross section of the pump volume, and an intensity ratio of about 8000 into the same solid angle between light emitted in the backward direction and orthogonal to it. 

Several other groups are working on various aspects of air lasing with molecular nitrogen or oxygen. A comparative study \cite{Ni_2012_AirLasingStudy} explored nitrogen lasing in a selection of gas mixtures (nitrogen with argon, neon or xenon). Several variations on the method were proposed by researchers who were interested in methods for the detection of trace impurities in the atmosphere and refer to it as standoff spectroscopy (SOS) \cite{Scully_2012_AirLasingProposal}. An interesting idea is to make use of the dispersion properties of light in air to overlap laser pulses of different wavelength at an arbitrary distance. Light pulses with smaller group velocity are followed with delayed pulses with larger group velocity. When the pulses overlap nonlinear processes including atmospheric self-focusing create a weakly ionized seed plasma filament, this plasma is then ``heated'' by another longer (ns) pulsed laser beam  to increase the plasma density, resulting in a stronger counter-propagating laser signal \cite{Sprangle_2011_HeaterAirLasing}. Another variation on the optical dissociation technique was demonstrated with water vapor \cite{Dogariu2021HydrogenLasing}. The authors showed backward and forward lasing from water vapor in air with a three-photon process: an initial photon dissociates hydrogen from oxygen, followed by two-photon pumping of hydrogen producing lasing on the Balmer-alpha line.

Apart from molecules present in the air (O$_2$, N$_2$ or water vapor), lasing activity can also be generated using existing atomic species in the atmosphere. For example, backward air lasing using argon with a molar fraction down to 10\% was demonstrated \cite{Dogariu2016ArgonLasing}. The authors have observed collimated, narrow-band and coherent emission at 1327\,nm after femto- and picosecond excitation at 261\,nm. Mechanisms of air lasing in argon have been recently elucidated in Ref.\,\cite{Nie2024}.

Stand-off atmospheric magnetometry can be envisioned by means of magnetic-field-dependent microwave scattering from laser-induced plasmas. The authors of Ref.\,\cite{Dogariu2019Plasma} have shown that the depolarization of plasma-scattered microwaves is affected by both the direction and magnitude of the magnetic field and could constitute a way to realize remote atmospheric vector magnetometry with nanosecond temporal resolution.

\section{Conclusions and outlook}

Remote-detection optical magnetometry is important for a variety of applications ranging from defense to basic atmospheric and geosciences. Although the field is more than half a century old, its development has not been as fast as in some other technological areas. However, the advent of powerful lasers and modern nonlinear-optics techniques gives the field a new boost, with ``sky magnetometers'' based on laser-guide star techniques appearing to be particularly promising.

% \begin{table}[htb]
%     \centering
%     \begin{tabular}{lccc}
%     \hline
%     \hline
%         Magnetometer & Typ. Sensitivity & Comment & Refs.  \\
%         \hline
%          CPT & & & \\
%          Remotely-interrogated vapor cell & & & \\
%          LGS-based & & & \\
%          Air lasing & & & \\
%          \hline
%          \hline
%     \end{tabular}
%     \caption{A summary table...}
%     \label{tab:my_label}
% \end{table}

Although this review is largely limited to Earth-based remote sensing, the discussed methods can also be applied in the field of space, and in particular, planetary magnetometry. Indeed, a general problem for spacecraft-based magnetometry is the familiar issue of discrimination of the external signals of interest from the platform noise \cite{Balogh-70699}. Here, remote detection may, once again, provide a powerful solution.   

We predict that the growing adoption of Artificial Intelligence (AI) will drive demand for advanced sensing techniques, particularly remote sensing. Historically, processing large, noisy datasets (for example, from urban environments \cite{dumont2022cities}) has been a major bottleneck. However, data processing automation, enabled by AI-driven correlation recognition and anomaly detection, now overcomes this obstacle \cite{Zhang2022}, fueling growth in the field reviewed here.

\section*{Acknowledgements}
The authors are grateful to Jon P. Davis, Arthur Dogariu, Paul Hickson, J\"urgen Matzka, Richard Miles, and Frank A. Narducci
%, and Stephen Potashnik 
for helpful discussions. This work was supported in part by the German Federal Ministry of Education and Research (BMBF) within the Quantentechnologien program (FKZ 13N15064 and FKZ 13N16457). FPB received support from the European Union’s Horizon 2020 research and innovation program under the Marie Sk\l{}odowska-Curie grant agreement No. 893150.

% %\appendix

%% \section{Units and conversion?}

% \DB{We will decide later if we need this or just give the conversions in the text}

% Magnetic field units: G, T, gamma

% Units concentration: at/cm$^3$, amg

% \section{Abbreviations?}

% CPT -- 

% NIST --

% ...

 \bibliographystyle{elsarticle-num} 
 \bibliography{report}





% \end{thebibliography}
\end{document}